\documentclass[journal]{vgtc}                     

\ifpdf
  \pdfoutput=1\relax
  \usepackage{graphicx}
  \DeclareGraphicsExtensions{.pdf,.png,.jpg,.jpeg}
\else
  \usepackage{graphicx}
  \DeclareGraphicsExtensions{.eps}
\fi

\graphicspath{{figures/}{images/}{./}}

\usepackage{microtype}
\PassOptionsToPackage{warn}{textcomp}
\usepackage{textcomp}
\usepackage{mathptmx}
\usepackage{times}

\usepackage{cite}
\usepackage{tabu}
\usepackage{comment}
\usepackage{changepage}
\usepackage{longtable}
\usepackage{pbox}
\usepackage{soul}
\usepackage{supertabular,booktabs,multirow,amsfonts,dcolumn}
\usepackage[table]{xcolor}
\usepackage{csquotes}
\usepackage{mdframed}
\usepackage[most]{tcolorbox}
\usepackage{colortbl}
\usepackage{threeparttable}
\usepackage[numbers]{natbib}
\usepackage{lscape}
\usepackage{tabularx}
\usepackage{xltabular}
\usepackage{placeins}
\usepackage{flushend}
\usepackage{tikz}
\usepackage{pgffor}
\usepackage{lipsum}
\usepackage{todonotes}
\usepackage{etoolbox}
\newcolumntype{P}[1]{>{\centering\arraybackslash}p{#1}}

\onlineid{0}
\vgtccategory{Research}
\vgtcpapertype{Evaluation} 

\title{From Code to Concept: Evaluating Multiple Coordinated Views in Introductory Programming}

\author{%
  \authororcid{Naaz Sibia}{0000-0001-7628-7077},
  \authororcid{Valeria Ramirez Osorio}{0009-0007-2320-7812},
  \authororcid{Jessica Wen}{},
  \authororcid{Rutwa Engineer}{0009-0003-4131-0996},\\
  \authororcid{Angela Zavaleta Bernuy}{0000-0002-1228-5774},
  \authororcid{Andrew Petersen}{0000-0003-1337-7985},
  \authororcid{Michael Liut}{0000-0003-2965-5302},
  \authororcid{Carolina Nobre}{0000-0002-2892-0509}
}

\authorfooter{
  \item Naaz Sibia, Jessica Wen, Rutwa Engineer, Andrew Petersen, Michael Liut, and Carolina Nobre are with the University of Toronto.\\
  E-mail: \{naaz.sibia, jessica.wen, rutwa.engineer, andrew.petersen, michael.liut\}@utoronto.ca, cnobre@cs.toronto.edu.
  \item Valeria Ramirez Osorio and Angela Zavaleta Bernuy are with McMaster University. E-mail: \{ramirev, zavaleta\}@mcmaster.ca.
}

\newcommand{\pvalue}[1]{%
  \ifdim#1pt>0.05pt%
    \textit{p} $={#1}$%
  \else%
    \ifdim#1pt<0.001pt%
        \textit{\textbf{p}} $\mathbf{<.001}$%
    \else
        \textit{\textbf{p}} $\mathbf{={#1}}$%
    \fi
  \fi
}

\newmdenv[
  backgroundcolor=gray!10,
  linecolor=gray!50,
  topline=false,
  bottomline=false,
  rightline=false,
  leftline=true,
  skipabove=\baselineskip,
  skipbelow=\baselineskip
]{takeawaybox}

\newcounter{quotecount}

\makeatletter
\newcommand\quotefontsize{%
  \@setfontsize\quotefont{8.8}{11}%
}
\makeatother

\newtcolorbox{myquotehelper}[1][]{
  breakable,
  colframe=gray,
  colback=white,
  sharp corners,
  left=10pt,
  right=0pt,
  top=0pt,
  bottom=0pt,
  boxrule=0pt,
  borderline west={1pt}{0pt}{gray},
  title={\scriptsize #1}
}

\long\def\comment#1{}

\abstract{
Novice programmers often struggle to understand how code executes and to form the abstract mental models necessary for effective problem-solving, challenges that are amplified in large, diverse introductory courses where students’ backgrounds, language proficiencies, and prior experiences vary widely. This study examines whether interactive, multi-representational visualizations, combining synchronized code views, memory diagrams, and conceptual analogies, can help manage cognitive load and foster engagement more effectively than single-visual or text-only approaches. Over a 12-week deployment in a high-enrollment introductory Python course ($N = 829$), students who relied solely on text-based explanations reported significantly higher immediate mental effort than those using visual aids, although overall cognitive load did not differ significantly among conditions. The multi-representational approach consistently yielded higher engagement than both single-visual and text-only methods. Usage logs indicated that learners’ interaction patterns varied with topic complexity, and predictive modeling suggested that early experiences of high cognitive load were associated with lower longer-term perceptions of clarity and helpfulness. Individual differences, including language proficiency and prior programming experience, moderated these patterns. By integrating multiple external representations with scaffolded support adapted to diverse learner profiles, our findings highlight design considerations for creating visualization tools that more effectively support novices learning to program.
}


\begin{document}
\maketitle

\section{Introduction}
\label{sec:intro}

Novice programmers often struggle to comprehend and trace code execution, especially when they lack prior experience in abstract reasoning about how variables, memory, and control flow interact \cite{sajaniemi2008study, murphy2012ability}. Although various visualization tools (e.g., Python Tutor) have been introduced to aid in understanding these details \cite{guo2013online}, beginners still frequently report feeling overwhelmed, grappling with overly concrete memory diagrams, and failing to generalize to higher-level conceptual models \cite{sorva2013review}. These challenges arise in part because novices need to \emph{bridge} abstract programming concepts and the concrete visualizations that illustrate runtime behavior. Merely showing memory layouts or step-by-step execution traces, while helpful for “seeing inside the black box” of code, may not alone foster the abstraction skills necessary for problem-solving \cite{sorva2013review, eckerdal2005novice}.

Educational research has long emphasized the value of Multiple External Representations (MERs), distinct but complementary formats that support different aspects of conceptual understanding \cite{ainsworth2006deft}. In mathematics and science domains, MERs are often scaffolded to guide learners from concrete depictions to more abstract generalizations, a strategy known as \emph{concreteness fading} \cite{nardi2014reflections, bettin2021frozen}. This coordination of representations can reduce cognitive load, enhance engagement, and facilitate transfer. 

To align with terminology in the visualization research community, we adopt the term \emph{Multiple Coordinated Views (MCVs)} \cite{scherr2008multiple} to refer to our implementation of synchronized code, memory, and analogy views. MCVs are a well-established idiom in VIS, typically used to support exploratory data analysis or linked insight across views. Our use of MCVs extends this idiom into educational settings, where learners interact with tightly linked views designed to foster conceptual understanding. Where we reference prior theoretical work from education, we retain the MER terminology; however, throughout this paper, we describe our system and visual scaffolding approach as an MCV design.

Despite the success of MERs and MCVs in their respective fields, programming education still lacks scalable, empirically evaluated tools that systematically integrate and coordinate these views. Most debugging or tracing tools rely on a single, detailed visual (e.g., stack frames) that may not adequately scaffold abstraction for novice learners \cite{sorva2013review}.

Furthermore, how different visualization approaches affect cognitive load, engagement, and performance remains a critical question. Some studies suggest that interactive diagrams can reduce mental effort \cite{de2017attention}, while others show that too much visual detail may overwhelm learners \cite{ainsworth2006deft}. Engagement, essential for persistence in computing, may depend on the type of visual scaffolding, the learner’s background, and the complexity of topics covered \cite{cunningham2019novice}. Language barriers and prior programming experience may also mediate how students perceive and benefit from visualizations \cite{guo2018non}.

\begin{figure*}
    \centering
    \setlength{\fboxsep}{3pt}
    \begin{minipage}{0.44\textwidth}
        \fbox{\includegraphics[width=\textwidth, height=4cm]{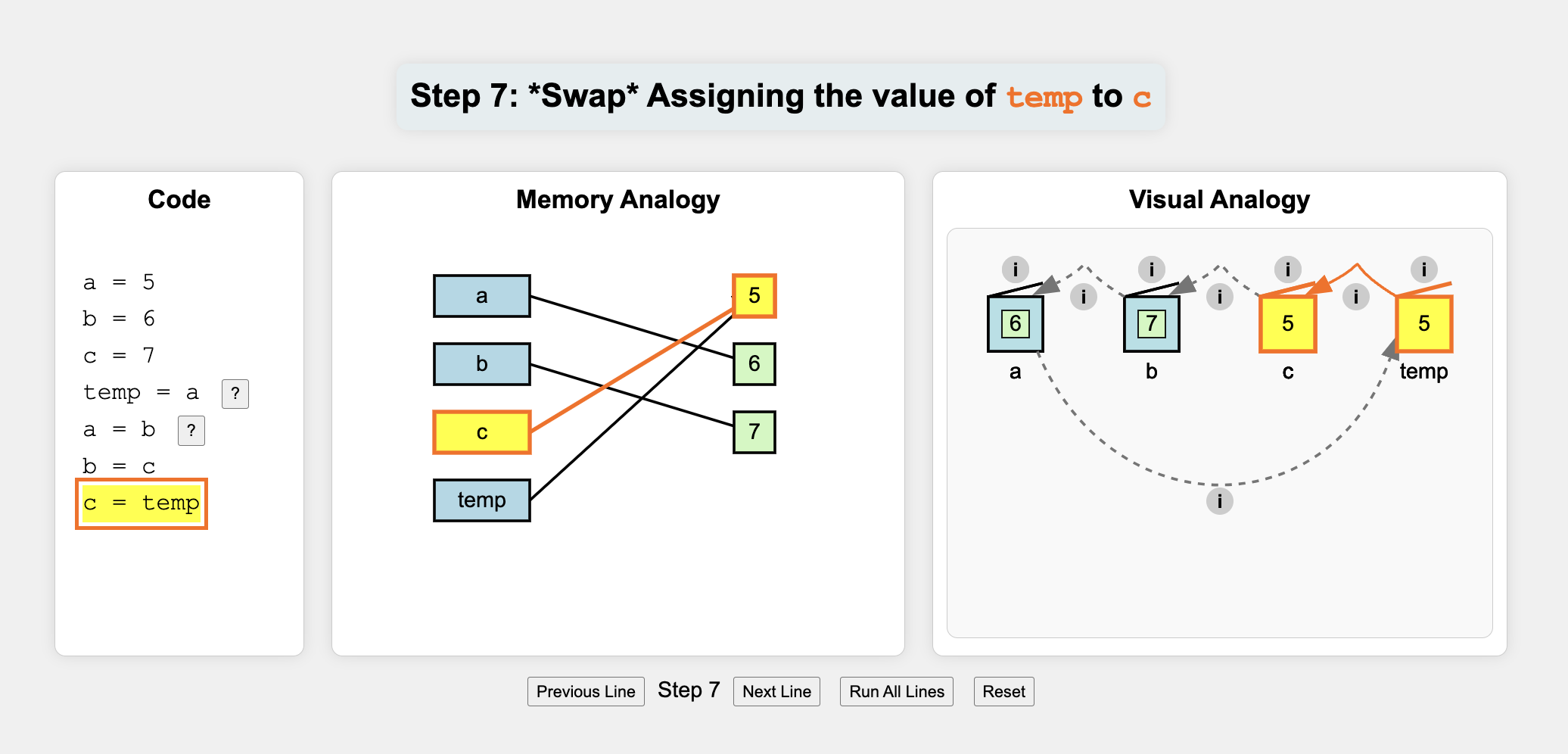}}
    \end{minipage}
    \hspace{5pt}
    \begin{minipage}{0.4\textwidth}
        \fbox{\includegraphics[width=\textwidth, height=4cm]{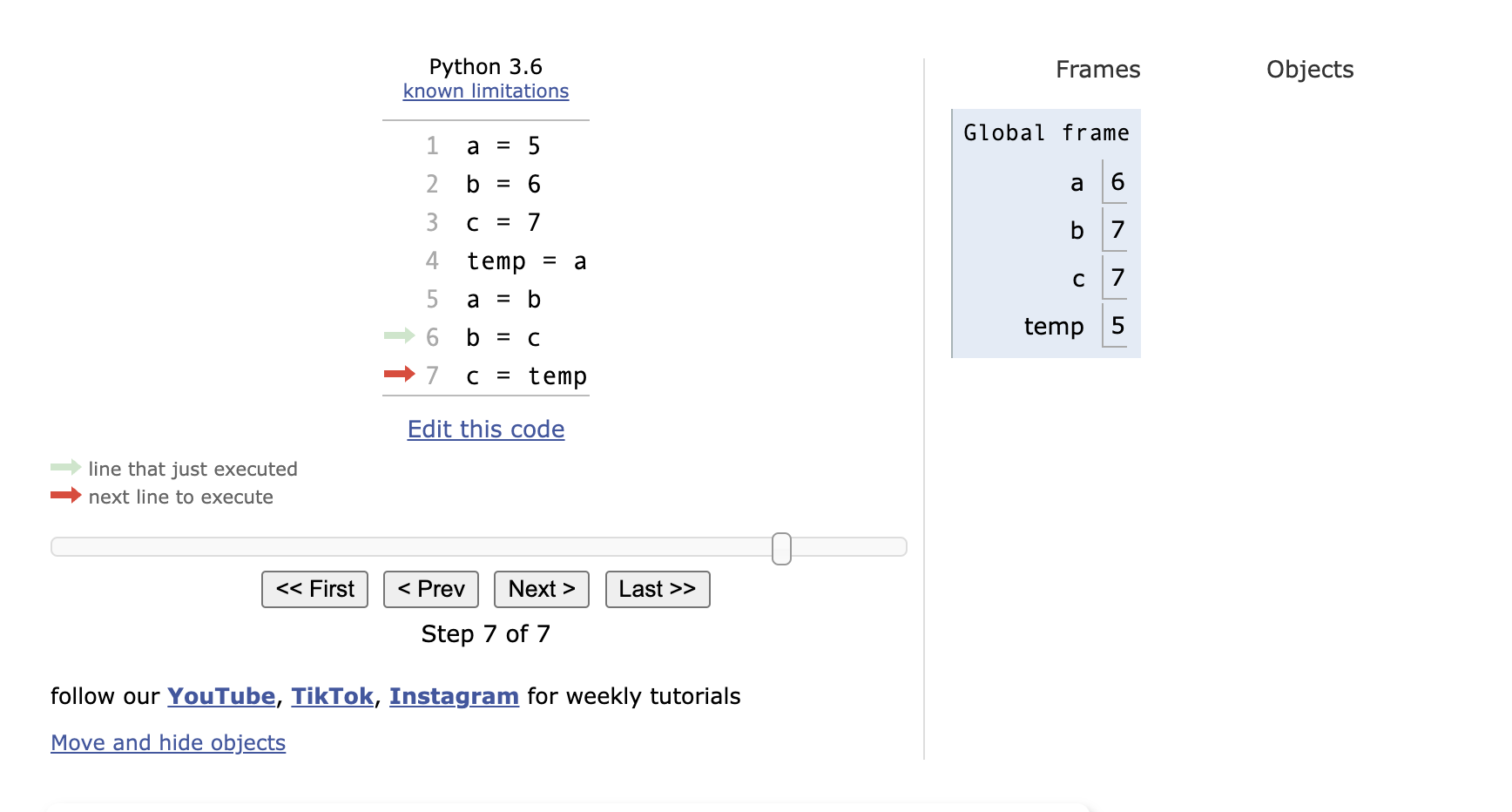}}
    \end{minipage}
    \caption{Left: The Multiple Coordinated Views (MCV) tool integrating synchronized code, memory diagrams, and abstract analogies in a step-by-step trace. 
    Right: Python Tutor (Single Visual condition), which provides a step-by-step execution trace with memory visualization.}
    \label{fig:visual-tools}
\end{figure*}

In response to these gaps, we conducted a 12-week field study in a high-enrollment introductory programming course at a research-intensive university in North America. We compared three conditions: (i)~a Multiple Coordinated Views (MCV) condition with synchronized code, memory, and analogy views; (ii)~a Single View condition using Python Tutor; and (iii)~a Text-Only condition with written step-by-step explanations. Figure~\ref{fig:visual-tools} illustrates these interfaces. We measured cognitive load, engagement, and usage patterns, and examined how learner characteristics, like prior experience and language proficiency, moderated these effects.

Specifically, our study addresses the following research questions:

\begin{itemize}
    \item[RQ1.] How do different representation types (Multiple Coordinated Views, Single Visual, Text Only) relate to novice programmers’ immediate mental effort, overall cognitive load, engagement, and performance in course assessments?
    \item[RQ2.] What learner- and task-related factors (e.g., prior experience, language barriers, complexity of code topics) moderate these effects, and how do they influence tool usage patterns over time?
    \item[RQ3.] In what ways do early struggles or successes with the assigned visualization condition shape students’ longer-term perceptions, including their final assessments of clarity and helpfulness?
\end{itemize}

The main contribution of this work is a large-scale, empirical evaluation of an MCV interface in an authentic educational setting. Our findings show that coordinated visual scaffolds can reduce immediate mental effort and promote engagement, particularly among subgroups with lower prior experience or greater language barriers. We also identify how early experiences shape long-term perceptions of clarity and usefulness, offering generalizable design heuristics for educational visualization systems.

The remainder of this paper is organized as follows. Section~\ref{sec:related} surveys prior research on visualization, MERs, MCVs, cognitive load, and educational tools. Section~\ref{sec:method} describes our study design and measures. Section~\ref{sec:results} presents the results, followed by implications for VIS design and computing education in Section~\ref{sec:discussion}.

\section{Related Work} \label{sec:related}

\subsection{Theoretical Underpinnings: Visualization, MERs, and Cognitive Load}

Research in the learning sciences has long demonstrated that \textit{visual representations} can reduce cognitive load and promote deeper conceptual understanding~\cite{heer2012interactive, mnguni2014theoretical}. In the visualization (VIS) community, scholars have similarly highlighted how external or interactive visual artifacts can amplify cognition by offloading memory demands and enhancing pattern recognition~\cite{card1999readings, munzner2014visualization, ware2012information}. According to Mnguni’s three-stage \textit{cognitive process of visualization}, learners internalize visual information, conceptualize it by connecting it to prior knowledge, and then externalize their mental models via sketches or other outputs~\cite{mnguni2014theoretical}. These stages underscore the importance of \textit{active engagement}—for instance, generating personal annotations—and \textit{concept formation}, such as recognizing meaningful patterns~\cite{mnguni2014theoretical}.

Moving beyond single, static diagrams, learners often benefit from \textit{multiple external representations} (MERs). Ainsworth’s DeFT framework (Design, Functions, Tasks)~\cite{ainsworth2006deft} posits that MERs can enhance understanding by (1) offering complementary perspectives of a concept, (2) constraining misinterpretation through connections between abstract and concrete views, and (3) promoting constructive learning through representational transformations. In mathematics education, Nardi~\cite{nardi2014reflections} has similarly argued for treating visual thinking as a primary language, consistent with Paivio’s Dual Coding Theory~\cite{paivio1991dual}, which highlights the cognitive advantage of coordinating verbal (e.g., source code) and non-verbal (e.g., graphical) information.

However, \textit{excessive} visual detail can overwhelm novices \cite{mayer2008increased}. Cognitive Load Theory (CLT) emphasizes balancing germane load (the cognitive effort that fosters learning) against extraneous load (the effort that hinders it)~\cite{paas2014cognitive}. Techniques such as \textit{concreteness fading} and progressive abstraction help guide learners from concrete, detailed depictions to more abstract representations~\cite{fyfe2014concreteness, munzner2014visualization}. For instance, Kordaki’s LECGO system~\cite{kordaki2010drawing} illustrates how learners can shift from intuitive drawings to increasingly formal text-based code, reducing extraneous cognitive load by anchoring abstract concepts in initially concrete forms. Similarly, Rubio et al.~\cite{rubio2015closing} demonstrate that \textit{physical computing}—providing tangible, real-world representations of loops, variables, and conditionals—can further engage novices (especially women) by reducing perceived difficulty and enabling more embodied interaction with program concepts.

\begin{takeawaybox}
\paragraph{Takeaway.} 
These theoretical perspectives (MERs, Dual Coding, CLT) underscore that well-designed visualizations should not only reduce extraneous cognitive load but also \textit{actively} foster concept formation. Incorporating multiple representations with careful scaffolding is key to cultivating robust, flexible mental models~\cite{hassan2024evaluating}.
\end{takeawaybox}

\subsection{Notional Machines, Mental Models, and Embodied Perception}

Within computing education, the principles outlined above underpin the concept of \textit{notional machines}—idealized models of program execution that guide novices’ developing mental models~\cite{sorva2013review}. These mental models enable students to simulate runtime behavior, reason about control flow, and debug more effectively~\cite{johnson1983mental}. Prior work~\cite{fincher2020notional, seel2017model} stresses that notional machines should help learners answer ``what if'' questions by providing faithful, if sometimes simplified, representations of computational processes. Whether these machines follow an \textit{Aristotelian} approach (striving for faithful abstraction) or a more \textit{Galilean} approach (deliberately distorting details to enhance comprehension) depends on the instructor’s pedagogical objectives~\cite{frigg2020models}.

Yet notional machines do more than merely depict hidden computational processes; they also shape students’ attentional and perceptual practices. Tenenberg~\cite{tenenberg2024notional} argues that these representational artifacts are inherently \textit{embodied} tools, influencing what learners notice and how they interpret what they see. This aligns with Adolph and Gibson’s~\cite{adolph2015gibson} view that learning involves discovering which variables in the environment are most relevant to the task at hand. In other words, notional machines and other representational devices should highlight the core elements of program behavior—such as loop conditions or object references—so that novices learn to ``see'' them as meaningful. However, as highlighted in a prior review~\cite{sorva2013review}, most commonly used visualizations in programming education focus on the \textit{how} of execution, treating memory states in granular detail. While this faithfully illustrates what the program is literally \textit{doing} in terms of manipulating objects and references, it offers little scaffolding for learners to grasp what the code is \textit{accomplishing conceptually} (e.g., the purpose of a loop or the invariant it maintains). For instance, Python Tutor provides precise memory traces but does not help learners abstract what the computation is \textit{about}~\cite{pollock2020essence}.

\begin{takeawaybox}
\paragraph{Takeaway.} 
By treating notional machines and MERs as \textit{perceptual} as well as cognitive tools, educators can direct novice programmers’ attention to the most conceptually significant features of code execution—shaping both how they think \textit{and} how they look at programs.
\end{takeawaybox}

\subsection{Visualization Tools for Programming: Empirical Evidence and Limitations}

A growing number of program-visualization tools aim to help novices ``see'' runtime behavior by showing step-by-step memory allocation and control flow (e.g., Python Tutor~\cite{guo2013online}). Although these tools effectively expose the hidden details of program execution, they often present highly granular views without systematically fostering higher-level abstractions~\cite{eckert2022loops, sorva2013review}. From a VIS perspective, such designs focus on revealing detailed data rather than guiding the user toward conceptual inferences~\cite{heer2012interactive, munzner2014visualization}. For instance, Python Tutor’s detailed frames and heap objects can unintentionally encourage novices to fixate on concrete outputs rather than conceptual structures—leading to misconceptions about loops~\cite{eckert2022loops}, arrays, or function behavior~\cite{karnalim2017effectiveness, karnalim2017use, balasubramanian2024challenges}.

Several empirical studies highlight the need to \textit{scaffold} the coordination of different representations. Kordaki’s multi-representational LECGO environment, for example, compares traditional code editors with a system that lets students alternate among drawing, pseudo-code, and C code~\cite{kordaki2010drawing}, revealing significant improvements in conceptual understanding. Similarly, Ma et al.~\cite{ma2011investigating} show that visualizations must go beyond static displays; they need to actively \textit{challenge} flawed mental models through cognitive conflict. Without explicit guidance on translating concrete memory traces into abstract notions (e.g., loop invariants, function contracts), novices revert to incomplete or incorrect understandings.

Rubio et al.~\cite{rubio2015closing} extend this point by illustrating how \textit{physical computing} can serve as another external representation, particularly effective for reducing the confidence gap among women in introductory courses. Their findings mirror the broader claim that engaging, multimodal experiences—whether through tangible hardware or progressive visual abstractions—can lower barriers to entry and improve retention.

Finally, Devathasan et al.~\cite{devathasan2022abstraction} demonstrate that \textit{abstraction proficiency} (i.e., learning to see the big picture rather than every memory detail) is strongly correlated with programming success ($r=0.70$). Students who struggle with abstraction often remain stuck in overly concrete depictions of code, reinforcing the importance of scaffolds that systematically transition from low-level details to conceptual views.

\begin{takeawaybox}
\paragraph{Takeaway.}
While existing tools (e.g., Python Tutor) are invaluable for exposing low-level execution, they frequently lack structured supports for \textit{representational fluency}: the ability to translate and integrate knowledge across code, memory, and conceptual abstractions. Evidence from Kordaki, Rubio, Eckert, and Devathasan suggests that multiple, carefully orchestrated representations---and explicit abstraction scaffolds---are critical to overcoming novices’ persistent misconceptions.
\end{takeawaybox}

\subsection{Positioning: Bridging the Gap through Abstraction-Focused Multiple Coordinated Views (MCVs)}
Although the learning sciences and mathematics education have extensively documented the value of \textit{multiple external representations}~\cite{ainsworth2006deft, nardi2014reflections}, computing instruction still lags in systematically integrating MERs. Most program-visualization tools privilege granular execution details (e.g., line-by-line traces) while overlooking the pedagogical need to help students \textit{abstract} away unessential details and interpret high-level patterns in code~\cite{sorva2013review, munzner2014visualization}.

Our work addresses this gap by explicitly designing a \textit{unified} environment that (1) leverages multiple visual representations of code to guide attention, (2) scaffolds the coordination of concrete and abstract representations, and (3) progressively fades from detailed views to conceptual models. In doing so, we build on insights from Kordaki’s drawing-based approach~\cite{kordaki2010drawing}, Rubio et al.’s findings on multimodal engagement~\cite{rubio2015closing}, Eckert et al.’s documentation of loop misconceptions~\cite{eckert2022loops}, and Devathasan et al.’s emphasis on abstraction~\cite{devathasan2022abstraction}.

By foregrounding \textit{abstraction} rather than execution tracing, we aim to help novices \textit{see} not just what the program does but what it \textit{means}. This study contributes new empirical evidence and design heuristics for designing effective Multiple Coordinated Views (MCVs) that support conceptual understanding in programming, a step toward educational visualizations that are both pedagogically grounded and aligned with visualization research principles.

\comment{
Notes:
\cite{adolph2015gibson} -- important to know which variables to pay attention to

\cite{greca2000mental} "... conceptual models are precise and complete representations that are coherent with scientifically accepted knowledge. That is, whereas mental models are internal, personal, idiosyncratic, incomplete, unstable and essentially functional, conceptual models are external representations that are shared by a given community, and have their coherence with the scientific knowledge of that community. These external representations can materialize as mathematical formulations, analogies, or as material artifacts."

\cite{seel2017model} "Ideally, the learner’s mental model will develop to also be able to answer these ‘what if’ questions by mentally simulating the processes or mechanisms being mastered."

\cite{ma2011investigating} the researchers identified that students held “‘non-viable’ mental models” and proposed and evaluated a method that used program visualization (reflecting an NM) and cognitive conflict to repair these models.

\cite{frigg2020models} - an Aristotelian model is still a true, albeit simplified, representation of the original system whereas a Galilean model typically is a more distorted representation of the system. 

\cite{fincher2020notional} 
- The purpose of idealization in NMs is to focus attention and to aid understanding. Omission helps focussing and inclusion helps making something comprehensible, for example by forming analogies.
- In general, NMs geared towards novices may add more “extra” than NMs geared towards advanced learners, or even experts.
}
\section{Method}
\label{sec:method}

\begin{figure*}[ht]
\centering
\begin{tikzpicture}[x=1.0cm, y=1.0cm, font=\footnotesize]
  \draw[->, thick] (0,0) -- (13,0) node[right]{Weeks};

  \foreach \x in {1,...,12} { \draw (\x,0.1) -- (\x,-0.1); }

  \node[rotate=45, anchor=west] at (1,0.3)  {Variables};
  \node[rotate=45, anchor=west] at (2,0.3) {Conditions};
  \node[rotate=45, anchor=west] at (3,0.3)  {Functions};
  \node[rotate=45, anchor=west] at (4,0.3) {For Loops};
  \node[rotate=45, anchor=west] at (5,0.3)  {Nested Loops};
  \node[rotate=45, anchor=west] at (6,0.3) {While Loops};
  \node[rotate=45, anchor=west] at (8,0.3) {Dictionaries};
  \node[rotate=45, anchor=west] at (10,0.3){Bubble Sort};

  \draw[fill=blue] (0.5,-0.7) circle (0.1) node[below=2pt]{Start Survey};
  \draw[fill=blue] (5,-0.7)   circle (0.1) node[below=2pt]{Midterm Survey};
  \draw[fill=blue] (6,-1.4)   circle (0.1) node[below=2pt]{Survey 6};
  \draw[fill=blue] (8,-1.0)   circle (0.1) node[below=2pt]{Survey 7};
  \draw[fill=blue] (10,-1.4)  circle (0.1) node[below=2pt]{Survey 8};
  \draw[fill=blue] (11,-1.0)  circle (0.1) node[below=2pt]{Final Survey};

  \draw[fill=red] (7,-0.3) ++(-0.075,-0.075) rectangle ++(0.15,0.15) node[below=4pt]{Term Test 1};
  \draw[fill=red] (11,-0.3) ++(-0.075,-0.075) rectangle ++(0.15,0.15) node[below=4pt]{Term Test 2};
  \draw[fill=red] (12.6,-0.3) ++(-0.075,-0.075) rectangle ++(0.15,0.15) node[below=4pt]{Final Exam};
\end{tikzpicture}
\caption{Course timeline: Weeks~1--8 show key topics; blue dots mark survey points; red dots mark assessments (Term Test~1 in Week~7, Term Test~2 in Week~11, Final Exam after Week~12).}
\label{fig:timeline}
\end{figure*}
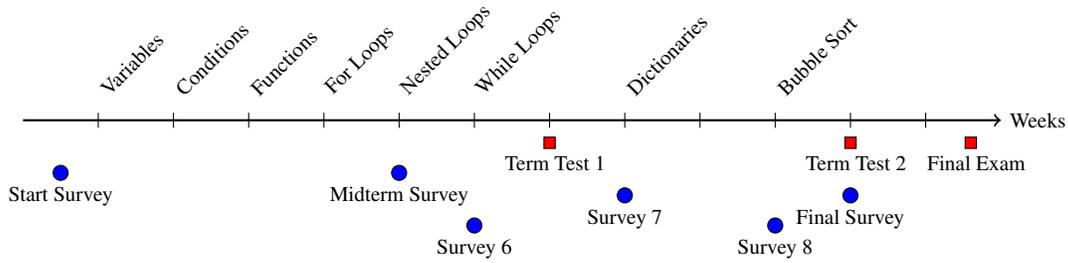

This section details the design, participants, materials, procedures, and analytical methods of our 12-week study, conducted in a large undergraduate introductory programming course at an R1 institution in North America. The study investigated the effects of three visualization approaches on novice programmers’ cognitive load, engagement, and performance in a real-world, field deployment setting.

\subsection{Study Context and Design}

A total of 1,080 students enrolled in the course, which introduced fundamental Python programming concepts over 12 weeks. Most students were first-year undergraduates, all learning the same content and possessing similar levels of initial programming knowledge. This homogeneity in background and curriculum supports a realistic field deployment and strengthens the ecological validity of our findings.

Of the enrolled students, those who consented to participate in the study ($N = 829$) were randomly assigned at the start of the term to one of three interface conditions: Multiple Coordinated Views (MCV), single view (SV), or text-only (TO). Participants remained in their assigned condition for the duration of the course (see Table~\ref{tab:experimental_conditions}). This between-subjects design enabled us to examine how varying levels of visual support influenced student learning and engagement across parallel weekly tasks.

We opted for a between-subjects design to maintain a consistent learning environment for each student and avoid potential carryover effects between interfaces, while still enabling comparison across the full cohort.

\begin{table*}[ht]
\centering
\caption{Summary of Experimental Conditions and Participant Counts}
\label{tab:experimental_conditions}
\resizebox{\textwidth}{!}{%
\begin{tabularx}{\textwidth}{@{}l c X@{}}
\toprule
\textbf{Condition} & \textbf{\# Students} & \textbf{Description} \\ 
\midrule
\textbf{Multiple Representations (MR)} & 271 & 
Interactive tool with synchronized views of source code, memory diagrams, and metaphorical analogies. Informed by MER theory~\cite{ainsworth2006deft} and cognitive load principles~\cite{paas2014cognitive}. Features dynamic linking, consistent color schemes, progressive disclosure, and interactive prompts. \\ 
\midrule
\textbf{Single Representation (SR)} & 270 & 
Step-by-step visualization of code execution and memory state using Python Tutor~\cite{guo2013online}. No additional abstraction layer. \\ 
\midrule
\textbf{Text-Only (TO)} & 270 & 
Worked examples in narrative form with step-by-step explanations. No diagrams or animations. \\ 
\bottomrule
\end{tabularx}
} 
\end{table*}
 

\subsection{Procedure and Timeline}

The study was embedded within the course’s flipped classroom structure \cite{zingaro2013facilitating}, where preparatory online homework was mandatory before attending each class. Students would watch videos about a concept at home, interact with the visual (or worked example text), and answer questions about it (depending on which group they were in). After this, students would solve homework questions about the concept. This ensured that students engaged with the visualizations early in the learning process, at a time when they were not yet accustomed to such tools. Data were collected through a series of online surveys integrated into the course requirements.

Table~\ref{tab:weekly_concepts} outlines the key programming concepts covered in the study. The Multiple Coordinated Views (MCV) condition used interactive visualizations, available at \href{https://github.com/CORE-Research-Lab/Interactive-Viz-IntroProgramming}{CORE Research Lab's GitHub repository}. The Single View (SV) group used Python Tutor (\href{https://pythontutor.com}{PythonTutor.com}), where students manually entered code to generate execution traces. The Text-Only (TO) condition received written step-by-step explanations hosted on OSF (\href{https://osf.io/cs9g7/?view_only=85bc2d687e4b4c51ab06616313113cdd}{OSF Repository}). In all conditions, students worked through weekly practice questions involving tracing memory behavior and variable updates, explaining code, and then extending the code to solve a new but related problem. These tasks were designed to target multiple levels of the Revised Bloom’s Taxonomy~\cite{krathwohl2002revision}, beginning with understanding and applying (e.g., tracing and interpreting code behavior), progressing to analyzing (e.g., explaining the function of key components), and culminating in creating (e.g., writing new code that builds on a prior solution). By including multiple question types in a sequence, the activity structure encouraged cognitive engagement across a hierarchy of reasoning processes, while also creating a consistent anchor point to evaluate how students interacted with different visual supports. Pre-condition analyses confirmed no significant differences in prior programming experience or language proficiency across groups.

\begin{table}[h]
\centering
\renewcommand{\arraystretch}{1.2}
\setlength{\tabcolsep}{6pt}
\caption{Concepts Covered Each Week and Corresponding Visualizations}
\label{tab:weekly_concepts}
\resizebox{0.48\textwidth}{!}{%
\begin{tabular}{ll ll}
\hline
\textbf{Week} & \textbf{Concept and Visualization} & \textbf{Week} & \textbf{Concept and Visualization} \\
\hline
Week 1 & \href{https://core-research-lab.github.io/Interactive-Viz-IntroProgramming/variable/}{Variables} 
       & Week 5 & \href{https://core-research-lab.github.io/Interactive-Viz-IntroProgramming/rainfall}{Nested Loops (Rainfall)} \\
Week 2 & \href{https://core-research-lab.github.io/Interactive-Viz-IntroProgramming/conditions/}{Conditions} 
       & Week 6 & \href{https://core-research-lab.github.io/Interactive-Viz-IntroProgramming/inspector}{While Loops (Inspector)} \\
Week 3 & \href{https://core-research-lab.github.io/Interactive-Viz-IntroProgramming/functions/}{Functions} 
       & Week 8 & \href{https://core-research-lab.github.io/Interactive-Viz-IntroProgramming/dictionary}{Dictionaries} \\
Week 4 & \href{https://core-research-lab.github.io/Interactive-Viz-IntroProgramming/simple-sum}{For Loops (Simple Sum)} 
       & Week 10 & \href{https://core-research-lab.github.io/Interactive-Viz-IntroProgramming/bubble-sorting/}{Bubble Sort} \\
\hline
\end{tabular}%
}
\end{table}

The study began in Week~1 with a start-of-term survey (S1) that collected demographic information, prior programming experience, and baseline confidence. To incentivize participation, students received a 2\% bonus on their final grade for completing both the start-of-term and end-of-term surveys. Intermediate surveys (S2--S5) were included as part of their preparatory homework in Weeks~5, 6, 8, and 10, ensuring near-universal participation as a course requirement. The final survey, administered in Week~12, gathered retrospective evaluations of tool usability, cognitive load, and overall effectiveness. Final course grades, computed on a 0--100 scale from homework, labs, midterm, final project, and final exam scores, were collected and linked to survey data using anonymized IDs.

\subsection{Visualization Design}
\label{sec:viz-design}

Our visualization designs build upon the theoretical frameworks outlined in Section~\ref{sec:related}. Specifically, we draw on Multiple External Representations (MER) theory \cite{ainsworth2006deft} to offer complementary views of program execution, while leveraging Cognitive Load Theory (CLT) \cite{paas2014cognitive} to minimize extraneous processing and foster germane load. We also integrate ideas from Dual Coding \cite{paivio1991dual} by combining verbal (source code) and non-verbal (diagrammatic) information in a synchronized manner. Below, we detail how these theoretical principles guided the features in each condition.

\subsubsection{Multiple Coordinated Views (MCV) Tool}

The \textbf{MCV tool} integrates three synchronized views to support a comprehensive understanding of program execution:
\begin{itemize}
    \item \textbf{Code Window:} Displays Python source code (verbal view), enabling learners to trace program logic step by step.
    \item \textbf{Memory Window:} Visualizes variables and runtime states through dynamic diagrams (non-verbal view), helping students see how data structures evolve in memory. This was similar to PythonTutor \cite{guo2013online}.
    \item \textbf{Analogy Window:} Presents instructor-developed metaphors (e.g., box-and-arrow or real-world analogies) that promote more abstract reasoning \cite{sanford2014metaphors}.
\end{itemize}

Consistent with MER theory \cite{ainsworth2006deft}, these three views offer complementary perspectives that learners can cross-reference. For example, a student can relate the memory diagram to the corresponding lines of source code and the high-level analogy, thereby reducing cognitive load by distributing information across multiple channels \cite{paivio1991dual}. We also employ dynamic linking \cite{ainsworth1999functions, roberts2006towards} (e.g., ``brushing and linking'') to highlight relevant elements across windows simultaneously, helping students build stronger connections between views. Additionally, progressive disclosure \cite{carroll1984training} mitigates extraneous load by revealing details only when needed, thus aligning with CLT principles of managing and fading complexity over time \cite{paas2014cognitive}.

\subsubsection{Single View (SV) Condition}

In the \textbf{SV condition}, we employed Python Tutor \cite{guo2013online} to offer a single, detailed visualization of code execution and memory state. While this tool effectively illustrates the concrete details of variable references and stack frames (a direct application of Dual Coding with verbal + diagrammatic information), it does not include a distinct abstract or metaphorical layer. From an MER standpoint, this means learners receive one primary view of runtime state without the additional vantage points that might support higher-level abstractions \cite{ainsworth2006deft}. Consequently, the SV approach may demand more effort from novices who need to mentally supply the abstract connections themselves.

We selected Python Tutor [17] as our ``single‐visualization'' baseline for several reasons. First, it is one of the most widely used free web‐based tools for Python tracing \cite{guo2013online}, with a user‐friendly interface that generates step‐by‐step memory diagrams. Second, unlike IDEs such as BlueJ \cite{kolling2003bluej} or jGRASP \cite{cross2004jgrasp} (commonly used for Java), Python Tutor is language‐specific to Python, making it directly relevant to our introductory course. Third, prior research \cite{kaila2010effects, karnalim2017use} shows that Python Tutor is already familiar to many novice learners, allowing a more ecologically valid comparison to our proposed multi‐views tool.

\subsubsection{Text-Only (TO) Condition}

The \textbf{TO condition} provides a set of carefully written, step-by-step worked examples \cite{muldner2022review} that explain code behavior entirely in plain language. In this design, learners rely exclusively on verbal processing, lacking the visual cues emphasized by Dual Coding \cite{paivio1991dual}. While text-based examples can be effective for some learners, the absence of diagrammatic or metaphorical views may result in higher cognitive load for novices who benefit from multiple, synchronized views of abstract code concepts.

\paragraph{Summary of Conditions.} The MCV design directly operationalizes MER and CLT by offering parallel, interconnected views and carefully managing extraneous details. The SV condition presents a single view (Python Tutor) that may reduce cognitive load compared to text alone but lacks higher-level analogies. Finally, the TO approach foregrounds verbal explanation and thus provides a baseline against which the benefits of visual scaffolding can be evaluated. 

A full walkthrough of the weekly instructional visuals used in the MCV condition, including their representational mappings and design rationales, is provided in Appendix~\ref{appendix:visuals}.

\subsubsection{Design Rationale and Alternatives}

While Section~\ref{sec:viz-design} describes the operational features of each condition, this section articulates the rationale behind our visual design decisions. Following established models of design rationale \cite{lee2020s}, we detail not only the features we implemented, but also the alternatives considered, constraints encountered, and trade-offs made in aligning visualization with pedagogical goals.

The design decisions behind PythonTutor are already justified by the authors in prior work \cite{guo2013online}. Here, we focus on the design rationale for the multiple view visuals that we designed.

\paragraph{Design Objectives.} Our primary goal was to support novice learners in forming coherent mental models of program behavior, particularly around control flow, memory manipulation, and abstract reasoning. To this end, we sought to integrate visual features that (i) reduce extraneous load through clarity and consistency, (ii) scaffold conceptual abstraction via metaphorical framing, and (iii) support representational fluency through cross-referenced views.

\paragraph{Design Alternatives.} Before finalizing the MCV interface, we explored several alternatives: (1) an animated character-based visualization (inspired by prior embodied tools such as \cite{rubio2015closing}), which we discarded due to visual clutter and interpretive ambiguity; (2) an analogy-only interface (without memory diagrams), which failed to anchor students' understanding in concrete execution details; and (3) static screenshots with color-coded highlights, which lacked interactivity and thus limited engagement. These explorations underscored the value of interactivity, visual coherence, and triangulation among views.

\paragraph{Mapping Views to Learner Goals.} We deliberately aligned each view with distinct cognitive functions: the code pane fosters verbal logic tracing, the memory pane anchors the execution state, and the analogy pane supports conceptual generalization. Early pilots revealed that novice learners often struggled to coordinate these views without explicit visual links. Thus, dynamic linking and synchronized highlighting were introduced to guide attention and reduce the demands on integration.

\paragraph{Iterative Refinement.} The MCV tool underwent three rounds of internal piloting with TAs and volunteer students. Key refinements included revising the visual view of variable states to enhance clarity, simplifying metaphorical text to reduce reading load, and calibrating the level of abstraction in analogies to prevent confusion. We also added progressive disclosure \cite{carroll1984training} to allow learners to explore complex traces at their own pace, consistent with concreteness fading principles.

\paragraph{Theoretical and Visual Trade-offs.} Balancing visual richness against cognitive overload was a recurring tension in the design process. While we aimed to highlight abstraction opportunities (e.g., loop patterns), we also had to avoid overwhelming novices with excessive details. This trade-off mirrors the tension between faithful design and comprehension oriented design observed in notional machines \cite{frigg2020models, tenenberg2024notional} and informed our decision to use simplified metaphors that emphasize \emph{why} code behaves a certain way, not just \emph{how}.


\subsection{Measures}
\label{sec:measures}

To evaluate how effectively our designs reflect the theoretical goals of reducing cognitive load and supporting active engagement (cf. Sections~\ref{sec:related} and \ref{sec:viz-design}), we employed the following instruments and data sources:

\paragraph{Cognitive Load.}
We measured students’ mental demand, effort, and frustration at multiple points (Weeks~5, 6, 7, 8, and~12) using a short version of the NASA-TLX \cite{hart1986nasa} and items adapted from Paas’s Cognitive Load Scale \cite{paas2014cognitive}. These scales align with CLT by capturing the extent to which each visualization condition introduces (or mitigates) extraneous load and supports germane load for concept formation.

\paragraph{Engagement.}
Engagement was assessed using items adapted from the Intrinsic Motivation Inventory \cite{brooke1996sus}, capturing enjoyment, interest, and perceived competence—factors critical to sustained attention and active engagement in MER contexts \cite{ainsworth2006deft}. The final survey (S6) included the System Usability Scale (SUS) \cite{brooke1996sus} for the MCV and SV conditions, while analogous text-based prompts were provided for TO. High engagement indicates that learners are effectively interacting with the views, as posited by Dual Coding and MER frameworks.

\paragraph{Code-Tracing and Open-Ended Responses.}
Intermediate surveys (S2--S5) included code-tracing tasks and open-ended questions. These tasks explore germane load by requiring students to process visual or textual clues, integrate them into a conceptual model of execution, and then apply that understanding to novel scenarios.

\paragraph{Usage Log Data (MCV and SV Conditions).}
For the MCV and SV tools, we recorded interaction metrics (time on task, step-through clicks, toggling between views, etc.). These usage logs reveal how learners navigate the multiple views or a single diagram, offering insight into the practical application of MER and CLT principles in a real learning environment (e.g., how often students switch between code, memory, and analogy, or backtrack on complex topics). In the absence of direct logging for TO, we rely primarily on self-reports and completion timestamps to approximate engagement.

\vspace{5pt}
\noindent
Together, these measures enable us to assess whether—and how—our theoretically grounded designs (Section~\ref{sec:viz-design}) translate into lower cognitive load and improved engagement in an authentic classroom setting.

\subsection{Data Analysis}

Quantitative data were analyzed using repeated-measures ANOVAs to compare changes in cognitive load and engagement across the six survey points and among the three conditions. When Mauchly’s test indicated sphericity violations, Greenhouse–Geisser corrections were applied. Significant effects were further examined using Tukey’s HSD or Bonferroni-corrected post-hoc tests. In addition, regression analyses and random forest models explored the predictive relationships between survey responses and students' final engagement and congitive load, with feature importance analyses identifying key determinants of student experiences. We also used $k$-means and hierarchical clustering on weekly survey response and logging data to identify changes in student engagement over time.

\paragraph{Qualitative Analysis.}
To enrich our interpretation of the quantitative findings, we conducted a reflexive thematic analysis of open-ended responses collected in the Week12 final survey. Students were invited to describe which aspects of the weekly visuals or worked examples they found helpful or challenging. Two researchers independently conducted open coding on a subset of responses and collaboratively developed a shared codebook. This was used to iteratively categorize responses around key themes such as perceived clarity, engagement, coordination demands, and affective reactions. The final coded dataset was analyzed for representational patterns and cross-condition differences. While not intended to yield generalizable results, this qualitative component provided interpretive depth and helped contextualize the measured effects of visualization type on learner experience (see Section~\ref{sec:results}).

\subsection{Ethical Considerations and Limitations}

All study activities were embedded within the course structure, with surveys constituting a required component of the online homework. Students were informed about the data collection and provided explicit consent, with a 2\% bonus on their final grade for completing the start-of-term and end-of-term surveys. Anonymity was maintained by linking survey responses to final grades via de-identified participant codes. Although random assignment by lab section ensured logistical feasibility, potential clustering effects and instructor influences could not be completely ruled out. Finally, while usage logs provided rich data for the MCV and SV conditions, engagement in the text-only condition was measured indirectly through time-on-task estimates.

In addition, our reliance on self-report instruments (e.g., NASA-TLX, IMI) introduces the possibility of social desirability bias or individual differences in interpreting survey items. The visualization materials were tailored for introductory Python, so our findings may not fully 
generalize to other programming languages or advanced topics. Our large and diverse sample enhances ecological validity but also brings variability in motivation and study habits that we could not fully control. Finally, while usage logs provided rich data for the MCV and SV conditions, engagement in the Text-Only condition was measured indirectly through time-on-task estimates, which may only approximate actual interaction behaviors.

\section{Results}
\label{sec:results}
We present our findings in three parts: 
\emph{(A)}~overall differences among the three representation types in terms of cognitive load and engagement, 
\emph{(B)}~weekly analyses based on cluster solutions from Weeks~6, 7, and~8, and 
\emph{(C)}~final survey outcomes combined with usage log data. 
Where applicable, we report statistical significance thresholds (e.g., $p < 0.05$), effect sizes, and relevant correlation coefficients.

\subsection{Effects of Representation Type on Cognitive Load \& Engagement}
\begin{figure}[h]
    \centering
    \includegraphics[width=0.45\textwidth]{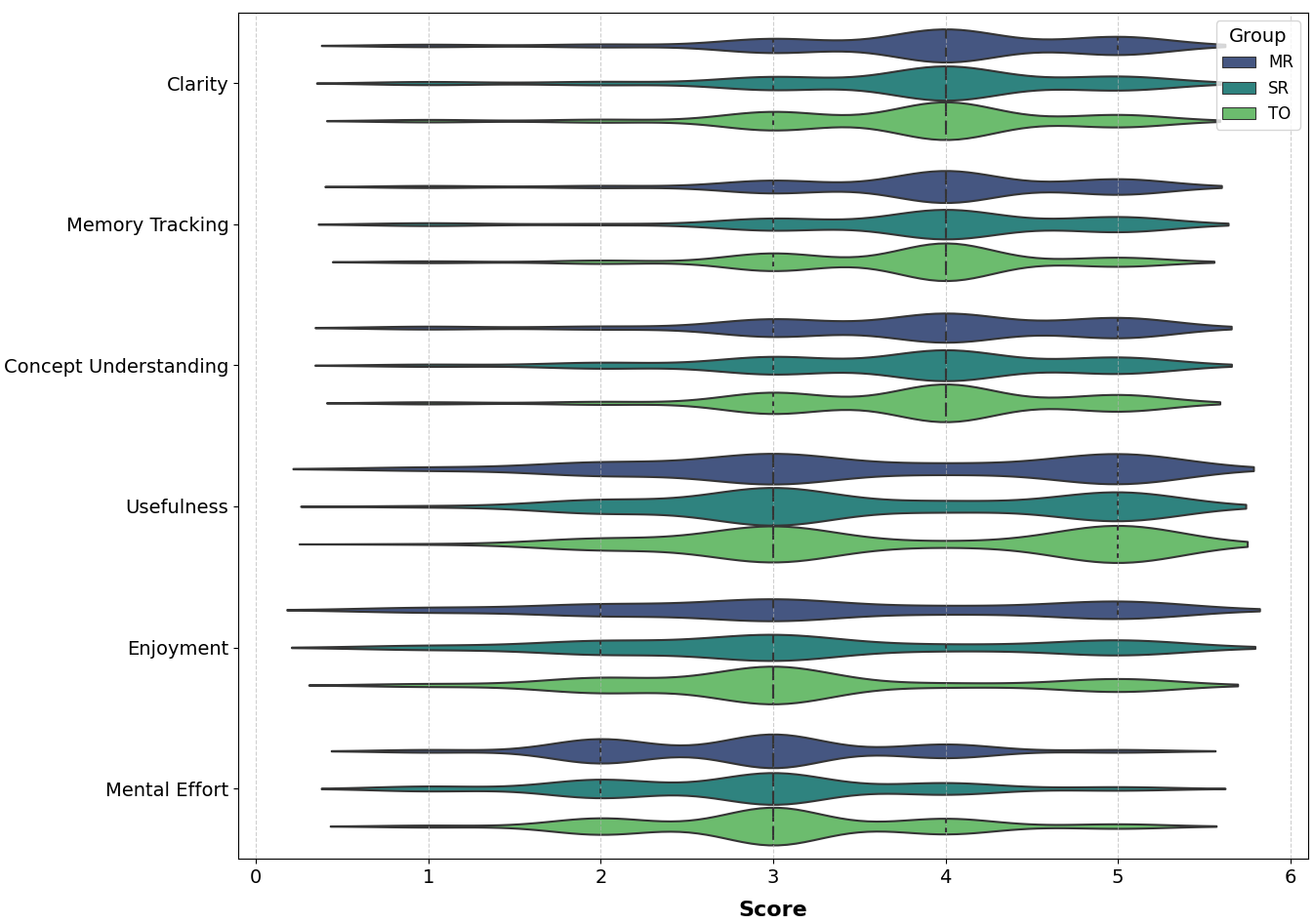}
    \caption{Differences in final cognitive load and engagement outcomes by representation type. Notes: The x‐axis denotes participants’ self‐reported mental effort on a 7‐point scale (1 = very low effort, 7 = very high effort). The y‐axis (violin width) indicates the distribution density of responses. We also overlay engagement scores, measured on a 5‐point scale (1 = low, 5 = high).}
    \label{fig:final-outcomes}
\end{figure}

Figure~\ref{fig:final-outcomes} visualizes the distribution of final outcomes across the three representation types--Multiple Coordinated Views (MCV), Single Representation (SR), and Text-Only (TO). The violin plots reveal key differences in mental effort, clarity, enjoyment, and related engagement measures. 

Students in the Text-Only condition reported the highest mental effort, aligning with the hypothesis that lack of visual support increases task difficulty. Conversely, the Multiple Views group exhibited higher engagement and clarity scores, reinforcing the idea that synchronized visual supports enhance learning experiences. Notably, the Single Visual condition displayed a more varied distribution, suggesting that some students benefited from it while others found it insufficient for engagement and comprehension. 

The following sections delve deeper into these patterns through ANOVA and post-hoc comparisons to further examine how representation type influenced mental effort and engagement.

\paragraph{ANOVA on Mental Effort}
Repeated-measures ANOVA revealed a significant effect of representation type on mental effort ($p=0.0161$, $\eta^2 = 0.0102$, $f = 0.10$), indicating a small to moderate effect of visualization format on perceived task difficulty.
 
Post-hoc tests indicated that students in the Text-Only condition reported significantly higher mental effort than those in the Single Visual and Multiple Visual groups. 
This finding suggests that having at least one form of visual support can alleviate some aspects of immediate task difficulty.

\paragraph{Overall Cognitive Load}
In contrast to mental effort, overall cognitive load did not differ significantly across the three conditions ($p=0.1496$). 
Although Text-Only participants perceived higher on-task difficulty, this difference did not translate into a measurable increase in their global sense of workload over time.

\paragraph{Engagement Findings}
Final survey data showed that participants in the Multiple Views group reported higher average engagement levels compared to the other two conditions. An ANOVA revealed a statistically significant effect of condition on engagement ($p = 0.0024$, $\eta^2 = 0.06$), with post-hoc comparisons indicating that engagement in the MCV condition was significantly higher than in the Single Visual condition ($p = 0.0134$). Interestingly, no significant difference in engagement was observed between the MCV and Text-Only groups ($p = 0.241$), and engagement in the Single Visual condition was lower than both.

However, weekly cluster analyses (Weeks 6–8) did not reveal a significant association between engagement-based cluster membership and visualization condition (Chi-square tests, all $p > 0.05$). This suggests that while the MCV interface may have supported higher engagement overall by the end of the term, this pattern was not consistent across time and likely interacted with individual learner differences and context-specific factors.

Taken together, the results suggest that while synchronized visual representations can promote engagement for some students, their effects are nuanced and may depend on personal preferences, evolving confidence, and external course pressures. Effect size calculations indicate a moderate impact of condition on engagement in the final survey, but more variable engagement patterns during the term.

\paragraph{Post-Hoc Analyses}
Beyond mental effort and engagement, we examined whether perceived clarity, memory support, helpfulness, overall usefulness, and enjoyment of the visualizations differed by condition.

No significant differences were detected in overall usefulness ($p = 0.1638$, $\eta^2 = 0.0045$) or visualization clarity ($p = 0.2424$, $\eta^2 = 0.0035$). However, memory support ($p = 0.0498$, $\eta^2 = 0.0075$) and enjoyment ($p = 0.0966$, $\eta^2 = 0.0058$) showed small but non-negligible effects that did not survive correction for multiple comparisons.

Taken together, these results suggest that while visual aids reduced perceived mental effort and enhanced engagement, they did not consistently enhance (or detract from) students’ clarity or perceptions of the visualizations’ overall usefulness and enjoyment.

\subsection{Weekly Analyses (Weeks 6, 7, and 8)}

To investigate evolving patterns in cognitive load, engagement, and clarity, we applied KMeans clustering to students’ survey data at Weeks 6, 7, and 8. The clustering procedure identified distinct profiles based on engagement, clarity, enjoyment, and perceived workload, and we used PCA to visualize these groupings.

\paragraph{Week 6}
A five-cluster solution best represented the data, grouping students by levels of engagement, clarity, enjoyment, and reported workload. 
However, a Chi-square test showed no significant association between cluster membership and condition assignment ($p>0.05$). Inspecting each cluster, it appeared that while some students within each condition felt well supported, others remained ambivalent, underscoring that engagement and clarity cannot be fully explained by the type of visualization alone.

\paragraph{Week 8}
We observed a four-cluster solution. 
Similar to Week~6, no strong link between condition and cluster membership emerged. 
Although the Multiple Visual group showed slightly higher mean engagement scores, the gap was not as pronounced as the overall ANOVA might suggest, suggesting individual differences in tool usage and preference at this point in the course.

\paragraph{Week 10}
Week~10 produced a seven-cluster solution, indicating a more fragmented set of student profiles. 
Despite this proliferation of clusters, no statistically significant relationship to experimental condition was identified. 

Together, the weekly analyses indicate that while students formed distinct motivational and cognitive load profiles, these did not map cleanly onto the three visualization conditions. Instead, the results highlight that condition alone cannot explain evolving learner experiences; individual characteristics (such as prior experience, language background, or coping strategies) and contextual pressures (e.g., assignment deadlines) play a substantial role. In other words, the clustering reinforces the need to interpret visualization effects through the lens of individual learner trajectories, rather than assuming uniform benefits across a group.

\subsection{Final Survey \& Usage Log Insights}

\paragraph{Correlation Analysis of Cognitive Load \& Final Perceptions}
The final survey, administered in Week~12, asked participants to reflect on their overall experience, including how helpful they found their respective visualization and how much mental effort they required. 
Correlation analyses revealed that higher initial cognitive load (measured in earlier weeks) correlated with lower later visualization clarity ($r=-0.2456,\,p<0.001$), higher mental effort ($r=0.2938,\,p<0.0001$), and lower enjoyment ($r=-0.2107,\,p<0.001$). 
These patterns suggest that students struggling early in the course might develop negative perceptions of the tool---regardless of its representation type---if they do not receive additional support or practice.

\begin{figure*}
    \centering
    \includegraphics[width=0.9\textwidth]{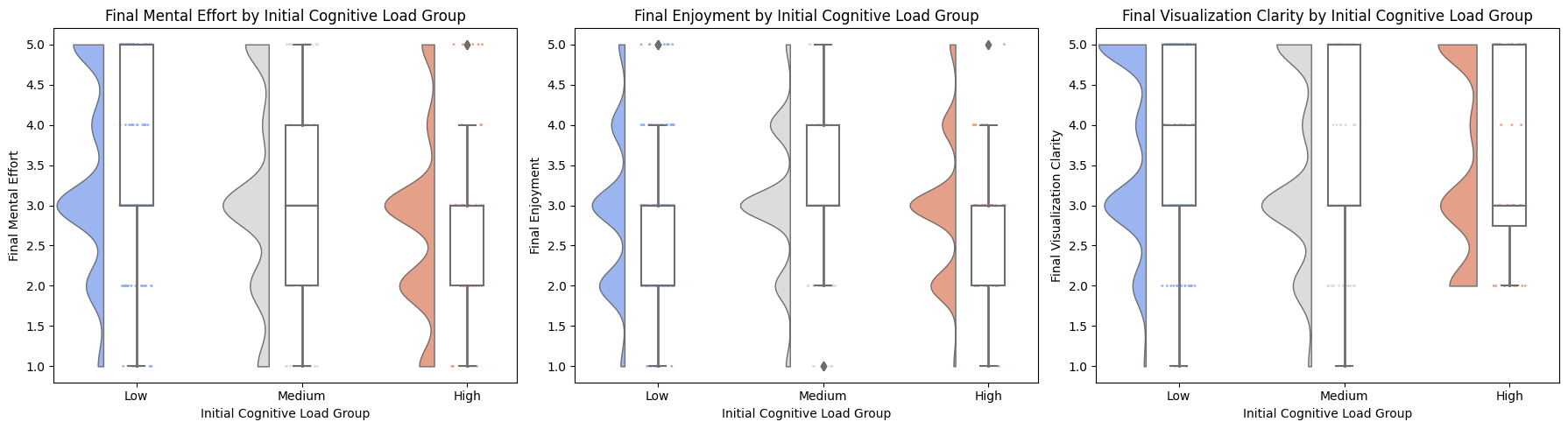}
    \caption{Effects of initial cognitive load on final mental effort, enjoyment, and visualization clarity. 
    Raincloud plots display distributions, quartiles, and individual responses across Low, Medium, and High initial cognitive load groups.}
    \label{fig:cognitive-load}
\end{figure*}

Figure~\ref{fig:cognitive-load} highlights how \textbf{students who began with high initial cognitive load} were more likely to report \textbf{greater final mental effort and lower enjoyment}. 
This aligns with the significant correlations observed, reinforcing the idea that early cognitive struggles may lead to persistently lower engagement with the visualization tool. 
Interestingly, final visualization clarity showed more variability across groups, indicating that while initial cognitive load was a predictor of later perceptions, it was not a definitive factor in students’ final clarity ratings.

\paragraph{Random Forest Model: Feature Importance for Engagement}
A random forest model, trained to predict end-of-term engagement, found that the strongest predictor was cognitive load (26.7\% importance). 
Other relevant factors included hours spent coding (8.3\%), difficulty understanding multiple sources (8.1\%), confidence in ability (7.3\%), and prior programming experience (7.1\%). 
Thus, while the choice of visualization matters, the data confirm that a variety of individual differences---such as hours spent coding, difficulty integrating multiple information sources, confidence in ability, and prior programming experience---shape students’ engagement trajectories.

\paragraph{Grade Performance \& Engagement}
Further statistical analyses (ANOVA and regression) showed that differences in perceived mental effort ($p=0.016$) and visualization clarity ($p=0.049$) were observable by condition, yet their direct relationship to final course grades was more nuanced. 
Although regression and LASSO models indicated that higher cognitive load negatively predicted visualization helpfulness and overall engagement, cluster membership tied to engagement patterns over time explained only a small portion of variance in final scores.

\paragraph{Post-Hoc Performance Analyses}
To further examine whether visualization condition affected conceptual understanding, we conducted ANOVAs on students’ performance by question across Test~1, Test~2, and the final exam. For each item, we report statistical significance ($p$), effect size ($\eta^2$), and Cohen’s $f$.

No statistically significant differences were observed across conditions on Test~1, suggesting that early-stage representational exposure may not have had an immediate impact on performance. However, several tasks on Test~2 and the final exam revealed small-to-moderate differences. Students in the Text-Only condition outperformed their peers on final exam items involving string manipulation ($p = 0.0139$, $\eta^2 = 0.0128$, $f = 0.114$), code tracing ($p = 0.0249$, $\eta^2 = 0.0113$, $f = 0.107$), and nested lists and loops ($p = 0.0209$, $\eta^2 = 0.0118$, $f = 0.109$). On Test~2, similar patterns emerged for unit testing ($p = 0.0166$, $\eta^2 = 0.0126$, $f = 0.113$) and tuple comprehension ($p = 0.0374$, $\eta^2 = 0.0094$, $f = 0.097$).

While most other questions yielded non-significant differences (see Table~\ref{tab:anova-results}), small effects were present in multiple areas ($f < 0.10$), suggesting that visual support may have uneven effects depending on the task. Somewhat counterintuitively, in a few cases the text-only condition slightly outperformed, which could be interpreted as evidence that requiring students to reason more deliberately without visual aids sometimes benefits performance. These findings underscore that representation type alone does not guarantee improved performance, and that other learner factors and task complexity likely modulate outcomes. Importantly, these quantitative patterns suggest that understanding students’ own strategies for working with or around the tools is critical to explaining why some conditions yielded surprising advantages. We therefore turn to the qualitative reflections in \nameref{sec:qual-results}, which shed light on how students constructed and adapted their own representations. 

\begin{table}[ht]
\centering
\small
\caption{ANOVA results for grade differences across groups by test and concept.}
\begin{tabular}{lrrrr}
\toprule
\textbf{Item} & \textbf{F-statistic} & \textbf{p-value} & \textbf{$\eta^{2}$} & \textbf{Cohen's $f$} \\
\midrule
\multicolumn{5}{l}{\textbf{Test 1}} \\
Variables (1 mark)              & 1.25 & 0.291 & 0.0044 & 0.067 \\
Conditionals (1 mark)           & 0.77 & 0.509 & 0.0028 & 0.053 \\
Nested For Loops (1 mark)       & 1.19 & 0.312 & 0.0042 & 0.065 \\
While Loops (1 mark)            & 1.56 & 0.198 & 0.0055 & 0.075 \\
List Methods (1 mark)           & 1.05 & 0.371 & 0.0037 & 0.061 \\
Scope (1 mark)                  & 0.80 & 0.494 & 0.0028 & 0.053 \\
Code Tracing (4 marks)          & 1.19 & 0.311 & 0.0042 & 0.065 \\
\midrule
\multicolumn{5}{l}{\textbf{Test 2}} \\
While Loops (1 mark)            & 2.98 & 0.031* & 0.0110 & 0.105 \\
Dictionaries (Q2, 1 mark)       & 1.40 & 0.240 & 0.0052 & 0.072 \\
Tuples (1 mark)                 & 2.83 & 0.037* & 0.0105 & 0.103 \\
Dictionaries (Q5, 1 mark)       & 2.37 & 0.070 & 0.0087 & 0.094 \\
Testing (1 mark)                & 3.44 & 0.017* & 0.0126 & 0.113 \\
Nested Lists (4 marks)          & 2.12 & 0.096 & 0.0078 & 0.089 \\
\midrule
\multicolumn{5}{l}{\textbf{Final Exam}} \\
Code Tracing (1 mark)           & 0.33 & 0.804 & 0.0012 & 0.035 \\
String Methods (1 mark)         & 3.56 & 0.014* & 0.0128 & 0.114 \\
Nested Lists (1 mark)           & 2.43 & 0.064 & 0.0088 & 0.094 \\
File I/O (1 mark)               & 0.97 & 0.409 & 0.0035 & 0.059 \\
\bottomrule
\end{tabular}
\begin{tablenotes}
\small
\item Note: * indicates $p < .05$. Effect sizes are reported as eta squared ($\eta^{2}$) and Cohen’s $f$.
\end{tablenotes}
\label{tab:anova-results}
\end{table}

\paragraph{Influence of Language Barriers}
When grouping students by the frequency with which they experience language barriers---Daily, Weekly, Monthly, and Never---we observed notable differences in visualization preferences. Students reporting language barriers preferred multiple visual representations, suggesting that the diversity of views might better support their comprehension ($p < 0.001$). In contrast, students who never experienced language barriers significantly preferred no visuals ($p < 0.001$). These findings underscore that language proficiency not only affects how learners interact with visual content but also suggests that tailored, supplemental textual scaffolding or language support may be essential for certain subgroups.

\paragraph{Usage Logs}
Analysis of clickstream data in the MCV and SV groups showed task-dependent interaction patterns. 
Simple operations on variables and conditionals averaged roughly three clicks, primarily stepping forward through code. 
Function-based tasks prompted more resets, possibly to re-examine parameter passing or return values. 
Loop constructs saw increased ``go back'' actions (around 12 clicks on average), while sorting or more complex algorithmic challenges sometimes exceeded 40 clicks, reflecting extensive backtracking. 
These differences suggest that topic complexity might influence how learners engage with the tools---and the need to repeatedly revisit specific steps---influences learners’ usage patterns and may modulate both perceived effort and engagement.

To further examine variations in user behavior, we conducted a K-means clustering analysis using event frequency data from sorting task interactions. The clustering was performed on a feature set comprising the number of times users performed key interactions such as `incrementLoop`, `decrementLoop`, `runAllIterations`, `resetLoop`, and `alert`.  The optimal number of clusters was determined using the elbow method and Silhouette analysis. The elbow method plots the within-cluster sum of squares (WCSS) against the number of clusters, revealing a diminishing return in variance explained beyond three clusters. This inflection point, or ``elbow,'' suggests that adding more clusters provides little additional explanatory power. 

Additionally, we computed the Silhouette score, which measures cluster cohesion and separation. A higher Silhouette score (closer to 1) indicates well-separated clusters with distinct behavioral patterns. The analysis revealed that three clusters achieved a balanced tradeoff between compactness and separation (\textbf{Silhouette score: 0.61}), outperforming alternative solutions with two or four clusters.

The identified clusters represent:
\begin{itemize}
    \item \textbf{High-Interaction Users:} Frequent backtracking and iterative refinement within loop-based operations.
    \item \textbf{Moderate-Interaction Users:} A mix of forward navigation and resets, balancing exploration with confidence in execution.
    \item \textbf{Low-Interaction Users:} Minimal revisitation of previous steps, possibly indicating either strong conceptual understanding or disengagement.
\end{itemize}

\begin{figure}[h]
    \centering
    \includegraphics[width=0.49\textwidth]{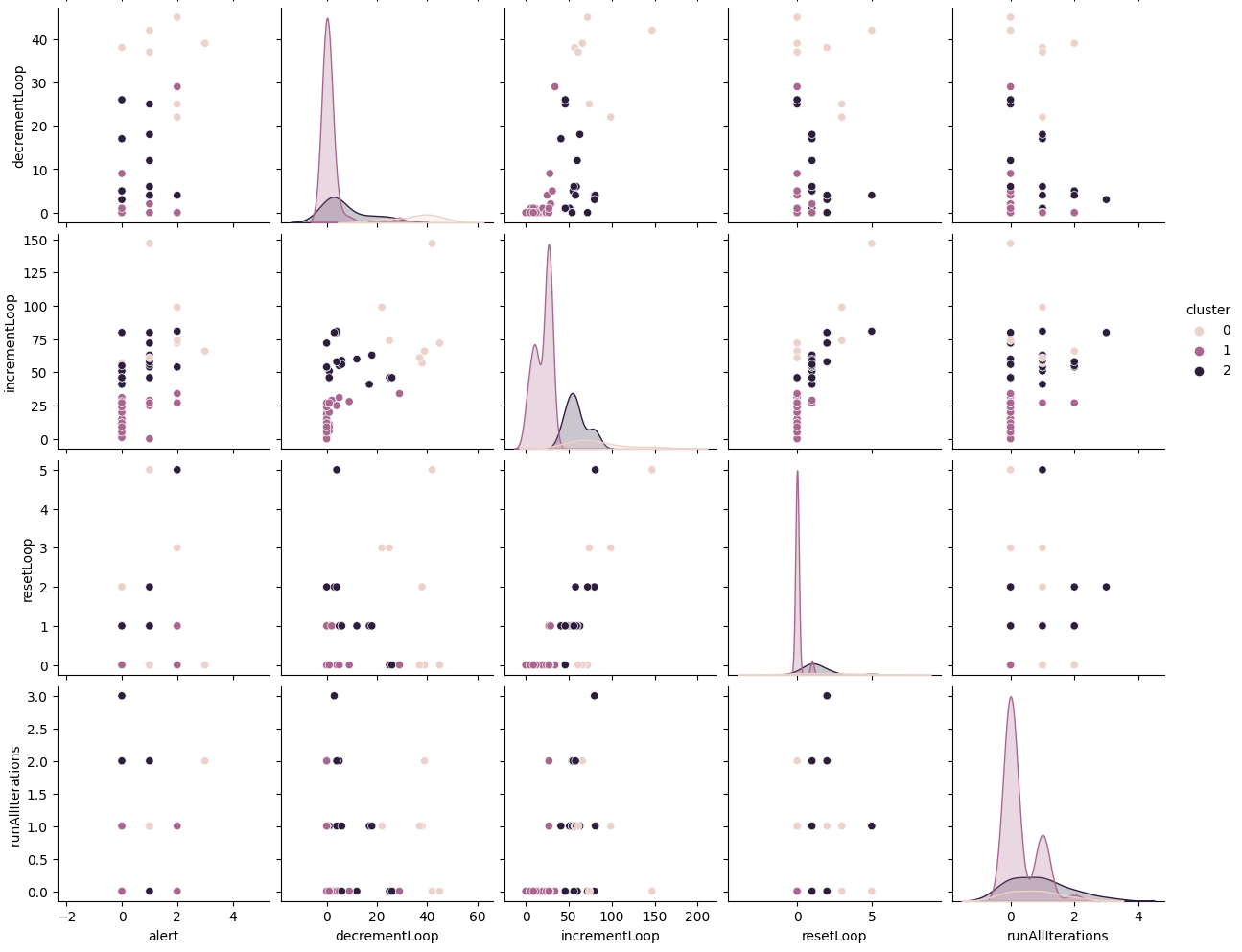}
    \caption{K-means clustering of user event frequencies for sorting tasks. The elbow method confirmed that three clusters provided the optimal balance of variance explained and model parsimony.}
    \label{fig:clustering}
\end{figure}

\subsubsection*{Qualitative Reflections}
\label{sec:qual-results}
The qualitative reflections help explain the patterns observed in the quantitative data, revealing how each representation type shaped both the experience of learning and, in some cases, performance outcomes.

In the Text-Only group, students consistently described the absence of visual scaffolds as making the material harder to follow---``It was hard to understand without any diagrams'', ``I had to draw it out by hand''. While this elevated the cognitive demands, many responded by generating their own diagrams and explanations. This self-construction appears to have been effortful in the moment but may have strengthened conceptual understanding for certain tasks, aligning with their stronger performance on items such as string manipulation, nested loops, and unit testing. In this sense, the extra difficulty sometimes acted as a productive constraint, prompting active processing and deeper encoding.

In the Single Visual condition, students often valued the clarity of step-by-step tracing---``Iterating through every step... helped me understand''---but also noted moments when the representation lacked explanatory depth---``Sometimes it just changed the memory and I wasn’t sure why''. The tool reduced the need to self-generate representations, which may have supported immediate comprehension, but the absence of stronger conceptual framing could explain why engagement and performance benefits were limited.

Students in the Multiple Views condition frequently highlighted how coordinated perspectives and analogies supported both engagement and comprehension---``It showed me what to look at'', ``The analogies helped make abstract ideas more concrete''. At the same time, several described an initial integration cost---``At first it felt like three tools, not one''. These accounts suggest that the multiple coordinated views created richer opportunities for connecting code to abstract concepts, but that the payoff depended on learners’ comfort in navigating and synthesizing multiple information sources.

Across all groups, the patterns point to a shared principle: representation type matters not only for its immediate cognitive effects, but for how it shapes the kind of work learners must do. Text-Only students bore the cost of self-construction, sometimes turning that into a performance advantage; Single Visual students followed execution with ease but risked missing conceptual integration; Multiple Views students engaged with a richer representational space, but success depended on representational fluency.
\section{Discussion}
\label{sec:discussion}

From a visualization research perspective, our study provides real-world evidence on how \emph{multi-view, educational visualizations} may be designed to support novices learning a complex, abstract domain. While our context is introductory programming, the design trade-offs we observed—between immediate cognitive relief and sustained conceptual growth, between concrete fidelity and abstract scaffolding, and between one-size-fits-all and adaptive pathways, may have relevance for other domains where learners integrate multiple conceptual and operational models. We position our three conditions not only as pedagogical interventions but also as contrasting \emph{visualization strategies}, allowing us to tentatively distill principles for when and how to combine multiple coordinated views for novice audiences.

Importantly, the term \emph{representation} has distinct meanings in the educational and visualization communities. In multiple external representations (MER) theory~\cite{ainsworth2006deft}, a representation can be textual, symbolic, diagrammatic, or metaphorical, each serving different cognitive functions. In visualization research, ``representation'' often refers to the encoding and mapping of data to visual marks, positions, and attributes. Our MR tool combined both senses: it integrated \emph{encoding-level variety} (syntax highlighting, spatial memory diagrams) with \emph{semantic variety} (abstract analogies) in a synchronized interface, enabling us to explore how these layers might interact for learners with diverse backgrounds.

\subsection{Reducing Immediate Cognitive Load Without Long-Term Gains}
In line with prior work suggesting that external diagrams can ease the immediate cognitive demands of code tracing \cite{kirsh2010thinking, cunningham2019novice}, our Text-Only group reported higher mental effort on short tracing tasks than both visual conditions. This finding aligns with the principle of \emph{reducing extraneous load} \cite{paas2014cognitive} through scaffolding. However, no significant differences emerged in overall cognitive load across the semester, which may indicate that learners eventually compensate through strategies such as self-drawn sketches \cite{cunningham2019novice}, note-taking or peer-explanations \cite{cohen2024factors, doebling2021patterns}.

\textbf{Design Implication}: This suggests that while visual scaffolds can lower the barrier to entry, their benefits may diminish unless paired with prompts for reflection or with strategies for gradually fading support. Prior work on scaffolding and concreteness fading suggests that intensive visual guidance can be particularly helpful for high-load introductory tasks, but should transition toward independent reasoning as learners’ expertise grows \cite{ainsworth2006deft}.

\subsection{Multiple Views Can Drives Engagement}
The Multiple Views (MR) condition yielded consistently higher engagement than the Single View and Text-Only conditions, supporting MER theory’s argument that multiple, coordinated representations offer diverse routes to understanding~\cite{ainsworth2006deft, paivio1991dual}. Here, the abstract analogy view appears to have added novelty and conceptual clarity beyond the shared memory diagram in both visual conditions, supporting the argument that analogies can be beneficial to learning programming \cite{bettin2021frozen}.

Interestingly, the Single View (Python Tutor) sometimes produced lower engagement than Text-Only, suggesting that visual fidelity alone is insufficient if learners cannot connect the details to higher-level meaning. Somewhat counterintuitively, this may occur because text forces sustained attention; students cannot simply glance at a diagram and move on, but must work through the reasoning step by step. In this sense, some of the ``gain'' for text-only conditions may reflect the productive/desirable difficulty of requiring effortful engagement \cite{bjork2020desirable}, rather than an inherent advantage of text over visualization.

\textbf{Design Implication:} In multi-view systems, complement concrete, faithful views with at least one abstract or metaphorical view that explicitly frames ``why this matters'' to sustain engagement.

\subsection{Extending MER Theory Through Interaction and Language Adaptation}
Our usage logs revealed that learners’ interaction patterns varied by both topic complexity (e.g., nested loops vs.\ sorting) and personal profile (e.g., prior experience, English proficiency). Students with lower English proficiency reported clearer benefits from MR, suggesting that synchronized multi-view designs can mitigate language barriers by shifting explanatory weight from text to visuals.

From a VIS perspective, this underscores the value of designing \emph{linguistically adaptive visualizations}, interfaces where text density, annotation language, or labeling style can adapt to user needs.

\textbf{Design Implication:} Integrate adaptivity not only by topic difficulty but also by \emph{linguistic accessibility}, allowing the same multi-view system to flexibly support multilingual or language-constrained users.

\subsection{Balancing Concrete Fidelity and Conceptual Abstraction}
Through a notional machines lens~\cite{tenenberg2024notional}, each tool tuned learners’ attention differently: Python Tutor’s faithful ``Aristotelian'' depiction of stack frames vs.\ the MR tool’s ``Galilean'' simplifications that foreground conceptual patterns~\cite{frigg2020models}. While detail-rich depictions serve precision, they risk cognitive overload if not paired with an abstract scaffold.

This mirrors concreteness fading~\cite{fyfe2014concreteness} strategies in other domains: beginning with concrete visuals, then progressively abstracting to more generic forms.

\textbf{Design Implication:} For novice-facing process visualizations, use \emph{progressive abstraction}: begin with concrete, faithful depictions and fade toward more abstract representations as fluency develops.

\subsection{Designing for Diverse Learner Trajectories}
Cluster analysis showed that students entering with high cognitive load often retained lower clarity and enjoyment, even in supportive conditions. Prior programming experience and English proficiency emerged as strong predictors of tool engagement.

For visualization designers, this reinforces the need for \emph{layered complexity}: allowing novices to toggle guided prompts, extended analogies, or simplified visuals on and off.

\textbf{Design Implication:} Offer \emph{adaptive layering} of visual and textual scaffolds, so mixed-ability audiences can self-regulate cognitive load.

\subsection{Relevance to Visualization Literature}
By embedding MER-based designs in an authentic, large-scale setting, we demonstrate how multi-view interfaces function not only as educational tools but also as laboratories for studying novice interaction with complex, dynamic systems. Our results refine MER theory by showing that:
(1) early cognitive relief must be coupled with scaffolding for sustained benefit,
(2) abstract views enhance engagement beyond what concrete fidelity offers, and
(3) adaptive, linguistically inclusive design is essential for broad accessibility.

These principles extend beyond programming to any domain where novices must bridge operational detail and conceptual models, e.g., biochemical pathways, or engineering simulations.

\subsection{Future Directions}
Future work should explore several avenues for advancing MER-based visualization design. One promising direction is \textbf{adaptive personalization}, in which visual complexity, pacing, and analogy style are tailored to individual learner profiles, using interaction logs to dynamically adjust support. Another is to investigate the \textbf{longitudinal impact} of such tools by tracking learners beyond a single course, examining whether initial engagement gains translate into durable conceptual change. Given our findings on language barriers, the development of \textbf{multilingual and multimodal tools}. For example, bilingual annotation layers, culturally relevant iconography, or audio–visual hybrids could reduce dependence on English proficiency and improve accessibility. Finally, further research should assess the \textbf{transferability of MER approaches to more complex topics}, such as recursion, concurrency, or advanced data structures, to test their scalability in higher-order domains.

In sum, our findings suggest that effective novice-facing visualizations require more than adding visual detail: they must integrate multiple views with complementary abstraction levels, adapt to linguistic and experiential diversity, and support a trajectory from early cognitive relief toward independent reasoning. These principles offer a roadmap for creating more adaptive, inclusive, and engaging visualization tools across domains.
\section{Conclusion}
\label{sec:conclusion}
This study examined how different visualization approaches--multiple views, a single-visual tool, and text-only--impact novice programmers’ cognitive load, engagement, and perceptions in a large introductory Python course. By embedding these tools in authentic homework tasks, we observed that multi-representational aids can reduce learners’ \emph{immediate} mental effort and foster higher engagement, although overall cognitive load across the semester did not differ significantly. Moreover, our analysis revealed how factors such as prior experience and language barriers moderate these effects, underscoring the need for adaptive scaffolds and multilingual support.

Ultimately, the effectiveness of visual tools in teaching programming hinges not only on what they display but on how they promote conceptual fluency, encourage early remediation of misconceptions, and accommodate diverse learner profiles. Our findings offer empirical guidance for designing inclusive, conceptually anchored visual aids that bridge the gap from code to deeper understanding—paving the way for more equitable outcomes in introductory programming courses.

\acknowledgments{
This work was supported by the Institute for Pandemics (IfP) Graduate Studentship Award, University of Toronto, and Natural Sciences and Engineering Research Council Discovery Grant PGS D – 600673 – 2025, \#RGPIN-2024-06005., and \#RGPIN-2024-04348
}

\bibliographystyle{abbrv-doi}
\bibliography{template}

\appendix
\section{Multiple Representation Visuals Used in Class}
\label{appendix:visuals}

This appendix documents the eight visualizations used in the classroom implementation of the MR tool. Each subsection includes a screenshot, learning objective, visual metaphor explanation, and rationale for design decisions, including trade-offs and refinement iterations.
gical intent, representational mappings, theoretical justification, and design trade-offs.

\subsection{Concept: Variable Reassignment and Swapping}

\begin{figure}[h]
  \centering
  \includegraphics[width=0.95\linewidth]{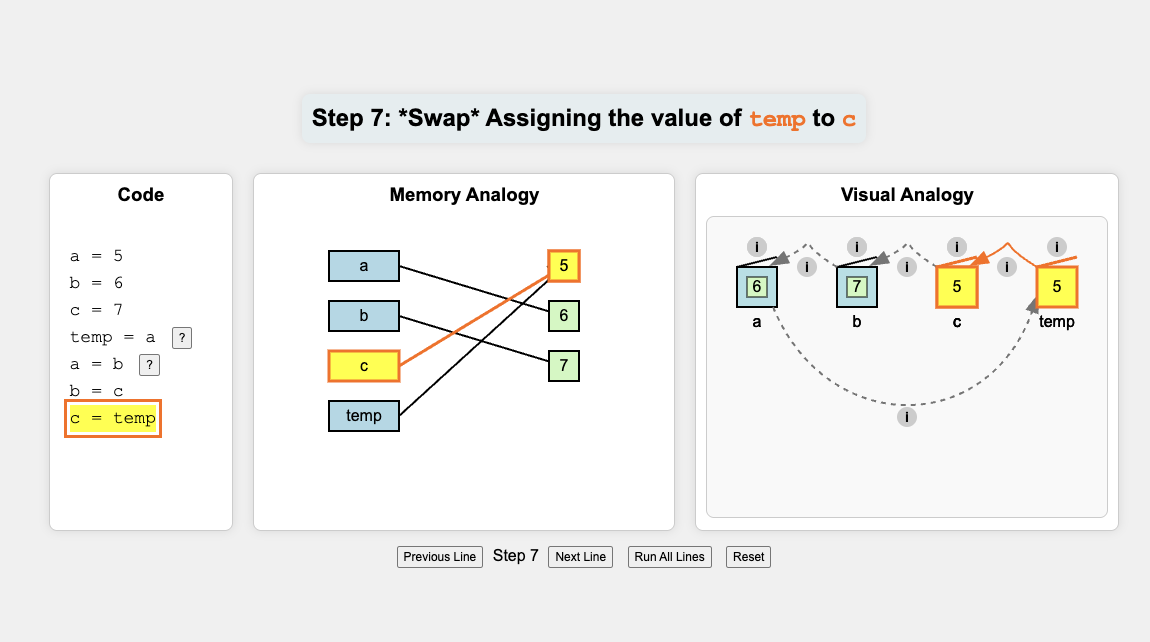}
  \caption{Step-by-step visualization of a variable swap using temporary storage.}
  \label{fig:swap-visual}
\end{figure}

\paragraph{Learning Objective.} To help learners understand how variables are reassigned during a multi-step swap operation, and how intermediate storage (i.e., a temporary variable) supports correct value transfer.

\paragraph{Representational Mapping.}
\begin{itemize}
    \item \textbf{Code Pane:} Highlights the active line \texttt{c = temp} to focus learner attention on the current assignment.
    \item \textbf{Memory Pane:} Visualizes current variable bindings and data flow through dynamic arrows and color cues.
    \item \textbf{Analogy Pane:} Offers a metaphorical depiction using arrows and containers to illustrate how values are passed between variables.
\end{itemize}

\paragraph{Design Justification.}
\begin{itemize}
    \item \textbf{Dynamic Highlighting:} Color coding synchronizes focus across all three panes, reducing extraneous load and supporting representational fluency.
    \item \textbf{Temporal Clarity:} Only the current step is shown in detail, consistent with progressive disclosure and split-attention principles.
    \item \textbf{Color Semantics:} Persistent colors (e.g., yellow for active values) reinforce identity and aid attention tracking.
    \item \textbf{Layout Coherence:} A left-to-right alignment of panes supports intuitive scanning from abstract to concrete representations.
\end{itemize}

\paragraph{Pedagogical Trade-offs.}
\begin{itemize}
    \item \textbf{Fidelity vs. Clarity:} Omitted memory models (e.g., stack frames) in favor of simplified, symbolic representations to reduce visual clutter.
    \item \textbf{Metaphor Calibration:} Visual analogies simplify control flow to support comprehension, with intentional abstraction from exact mechanics.
    \item \textbf{Cognitive Load Management:} Removed unnecessary details to avoid overwhelming novices, aligning with principles from Cognitive Load Theory and concreteness fading.
\end{itemize}

\noindent
This visualization reflects our broader strategy of scaffolding learner understanding through synchronized, complementary representations that reinforce the formation of coherent mental models.

\subsection{Concept: Function Calls and Scope}

\begin{figure}[h]
  \centering
  \includegraphics[width=0.95\linewidth]{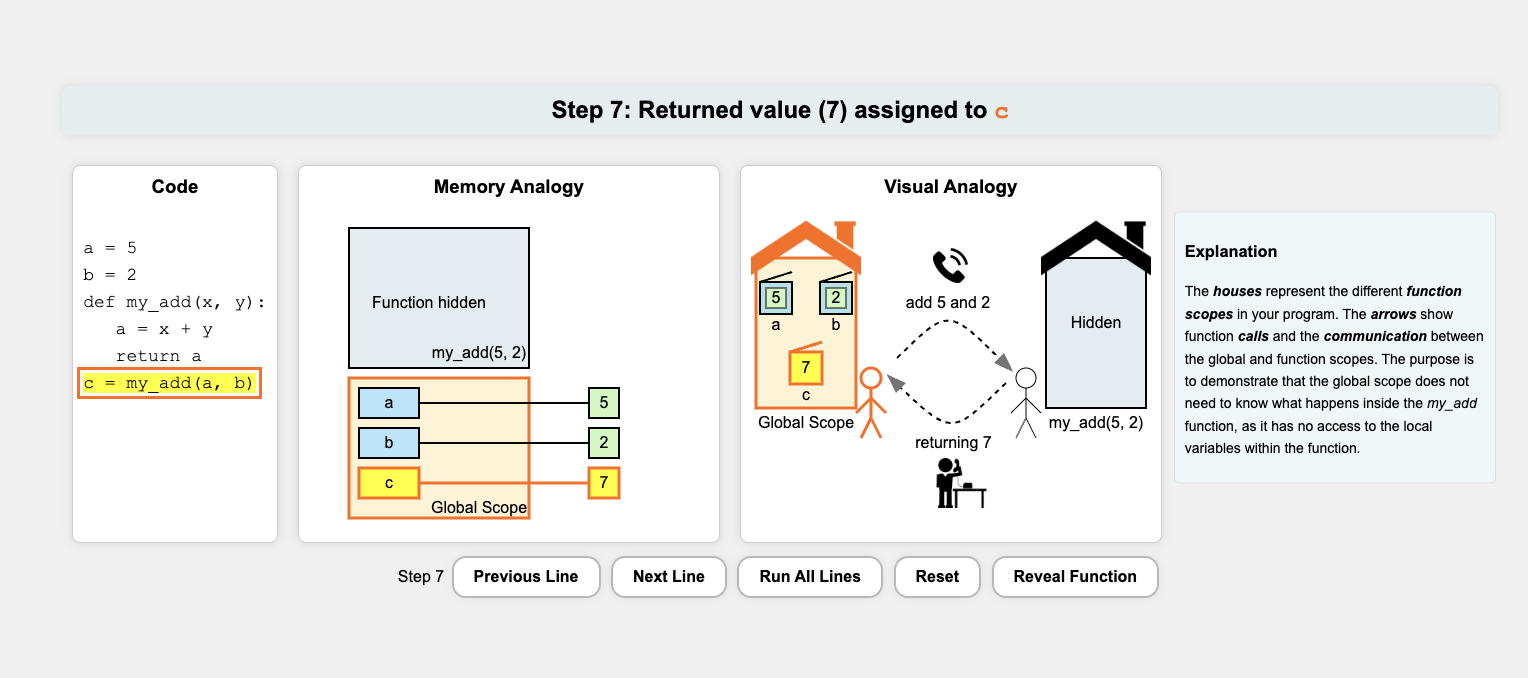}
  \caption{Visualization of a function call and value return between scopes.}
  \label{fig:function-call-visual}
\end{figure}

\paragraph{Learning Objective.}
To help learners distinguish between global and local variable scopes, understand how function arguments are passed, and how values are returned and assigned in the global context.

\paragraph{Representational Mapping.}
\begin{itemize}
    \item \textbf{Code Pane:} Highlights the line \texttt{c = my\_add(a, b)} to focus attention on the moment a function call is made and its return value is assigned.
    \item \textbf{Memory Pane:} Distinguishes between the global scope (visible) and the function scope (hidden), showing only accessible bindings.
    \item \textbf{Analogy Pane:} Uses visual metaphors of “houses” and “phone calls” to represent the separation of scopes and the communication between them.
\end{itemize}

\paragraph{Design Justification.}
\begin{itemize}
    \item \textbf{Scope Metaphor:} The house metaphor reinforces the concept of encapsulation and reinforces that global code cannot access local function internals.
    \item \textbf{Communication Arrows:} Dashed arrows visually depict the call-return process, helping learners conceptualize how parameters and return values are passed between contexts.
    \item \textbf{Dynamic Attention Cueing:} The active line and updated variable (\texttt{c = 7}) are highlighted using consistent color conventions (yellow/orange) across panes.
    \item \textbf{Cognitive Scaffolding:} The optional “Reveal Function” feature allows learners to gradually uncover internal function details—consistent with progressive disclosure principles.
\end{itemize}

\paragraph{Pedagogical Trade-offs.}
\begin{itemize}
    \item \textbf{Simplified Scope Model:} The internal execution of the function is abstracted away to reduce extraneous load and foreground the idea of black-box abstraction.
    \item \textbf{Use of Anthropomorphism:} The stick figures and houses create a personified model of data exchange that supports novice reasoning, though it abstracts away memory-level precision.
    \item \textbf{Focus on Return Mechanism:} Function internals are not animated step-by-step; instead, emphasis is placed on the output and variable assignment. This choice prioritizes *conceptual clarity* over procedural trace fidelity.
\end{itemize}

\noindent
This visualization supports abstraction by depicting a function as a separate, encapsulated process and helping students focus on scope boundaries, return values, and the idea of communication across program components.

\subsection{Concept: Conditional Branching and Control Flow}

\begin{figure}[h]
  \centering
  \includegraphics[width=0.95\linewidth]{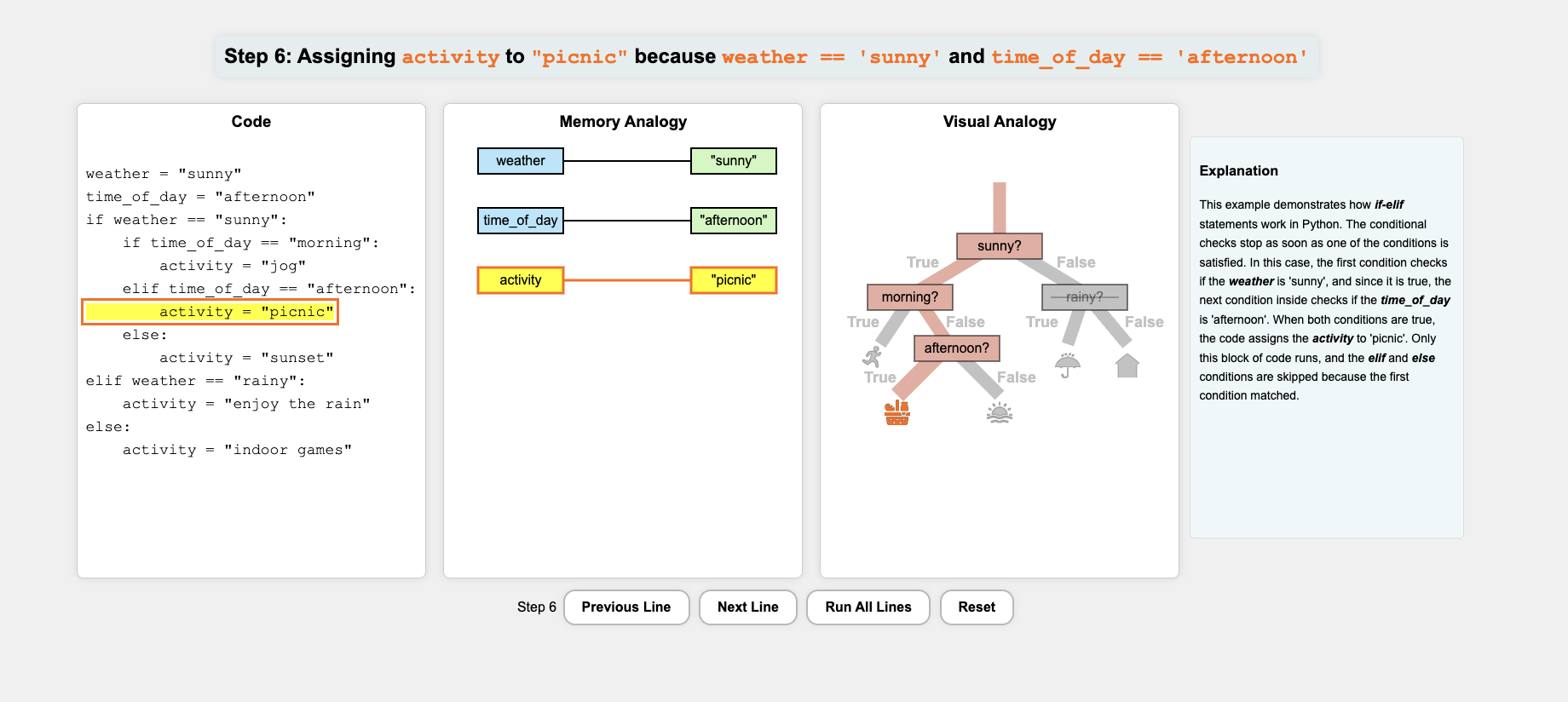}
  \caption{Visualization of nested \texttt{if-elif-else} statements and their effect on variable assignment.}
  \label{fig:conditionals-visual}
\end{figure}

\paragraph{Learning Objective.}
To support learners in understanding how nested conditionals evaluate sequentially, how short-circuiting occurs in \texttt{if-elif-else} structures, and how logical paths map to control flow outcomes.

\paragraph{Representational Mapping.}
\begin{itemize}
    \item \textbf{Code Pane:} Highlights the specific conditional block executed based on the current variable values (\texttt{weather}, \texttt{time\_of\_day}).
    \item \textbf{Memory Pane:} Shows current values of the condition variables and the result of the executed assignment (\texttt{activity = "picnic"}).
    \item \textbf{Analogy Pane:} Depicts a binary decision tree metaphor, where each node corresponds to a condition, and branches indicate logical paths (True/False).
\end{itemize}

\paragraph{Design Justification.}
\begin{itemize}
    \item \textbf{Tree Metaphor:} The decision-tree analogy visually encodes control flow decisions. The orange traversal path reinforces the logical evaluation process, while grayed-out branches help clarify what was not executed.
    \item \textbf{Outcome Anchoring:} The final result (“picnic”) is represented both as a memory value and as an endpoint in the tree, reinforcing causal reasoning.
    \item \textbf{Color Consistency:} Orange and yellow are used to maintain continuity in highlighting across panes, drawing learner attention to active state and logic.
\end{itemize}

\paragraph{Pedagogical Trade-offs.}
\begin{itemize}
    \item \textbf{Selective Path Highlighting:} Only the active branch is shown in full color to minimize extraneous load, but this may limit learners’ ability to compare all potential paths unless they revisit earlier steps.
    \item \textbf{Abstracted Evaluation Logic:} Conditions are presented visually as discrete checks rather than expressions to evaluate—this simplifies understanding, but removes granularity (e.g., logical operators).
\end{itemize}

\noindent
This visualization targets control flow comprehension by integrating literal (code), symbolic (memory), and conceptual (tree) views of conditional branching. It supports novices in tracing execution paths and understanding how conditionals govern assignment logic.

\subsection{Concept: Iteration and List Traversal}

\begin{figure}[h]
  \centering
  \includegraphics[width=0.95\linewidth]{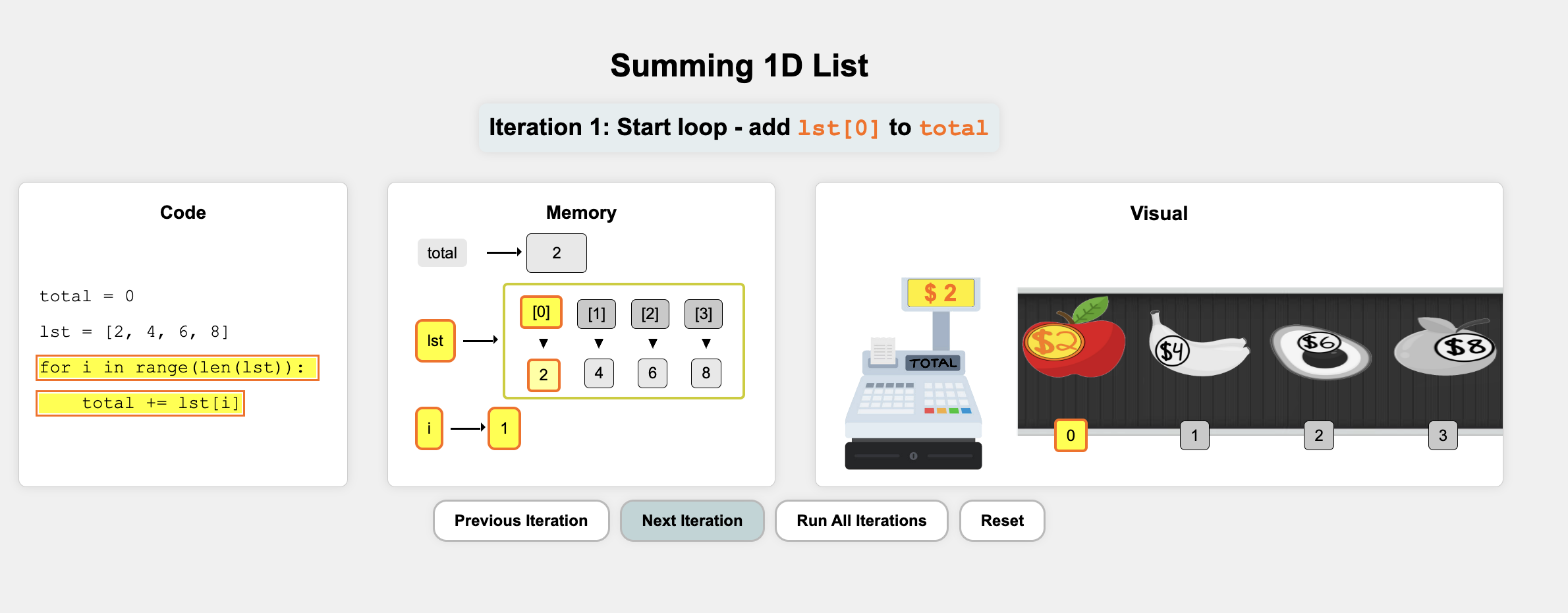}
  \caption{Visualization of a \texttt{for} loop traversing a list and accumulating values into a total.}
  \label{fig:loop-sum-visual}
\end{figure}

\paragraph{Learning Objective.}
To help novices understand the role of indexing in list traversal, loop iteration over array positions, and the cumulative update pattern using a running total.

\paragraph{Representational Mapping.}
\begin{itemize}
    \item \textbf{Code Pane:} Highlights each loop iteration and the accumulation operation (\texttt{total += lst[i]}).
    \item \textbf{Memory Pane:} Displays the loop index \texttt{i}, the current list item (\texttt{lst[i]}), and the evolving value of \texttt{total}.
    \item \textbf{Analogy Pane:} Uses a supermarket conveyor belt metaphor to visualize iteration over indexed items (e.g., fruits with prices), where the cash register represents the accumulating total.
\end{itemize}

\paragraph{Design Justification.}
\begin{itemize}
    \item \textbf{Conveyor Metaphor:} The moving belt with labeled prices maps clearly to indexed iteration, making each access concrete and visually distinct.
    \item \textbf{Running Total as Cash Register:} Abstract accumulation is represented as a price tally, reinforcing the mental model of iterative summing.
    \item \textbf{Synchronous Highlighting:} The current list index, list value, and total are simultaneously emphasized across panes to reduce split attention and clarify relationships.
\end{itemize}

\paragraph{Pedagogical Trade-offs.}
\begin{itemize}
    \item \textbf{Literal Indexing Focus:} This visualization emphasizes index-based access, which aligns well with traditional CS2 approaches but may obscure the benefits of idiomatic iteration (e.g., \texttt{for x in lst}).
    \item \textbf{Item Type Simplification:} Using grocery prices abstracts away from type heterogeneity or complex data structures, which is pedagogically useful early but may limit transfer.
\end{itemize}

\noindent
This visualization scaffolds the cognitive process of loop iteration, enabling learners to link iteration steps with memory access and accumulation in a narratively intuitive context.

\subsection{Concept: Nested Loops and Multi-Level Indexing}

\begin{figure}[h]
  \centering
  \includegraphics[width=0.95\linewidth]{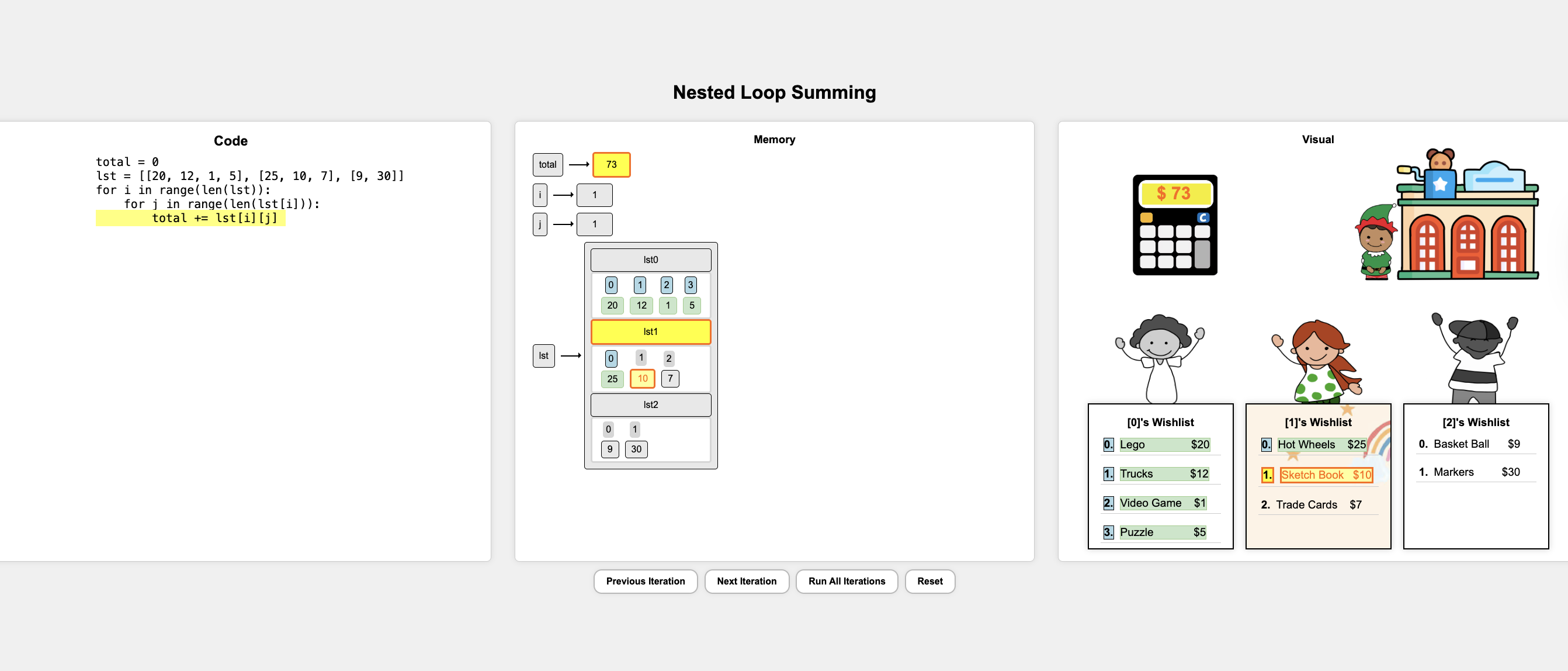}
  \caption{Visualization of a nested \texttt{for} loop summing values in a 2D list using \texttt{total += lst[i][j]}.}
  \label{fig:nested-loop-sum-visual}
\end{figure}

\paragraph{Learning Objective.}
To help students understand how nested loops traverse multi-dimensional data structures, and how two indices (\texttt{i}, \texttt{j}) interact to access inner elements within nested lists.

\paragraph{Representational Mapping.}
\begin{itemize}
    \item \textbf{Code Pane:} Highlights the nested loop structure and the active line where the sum operation occurs.
    \item \textbf{Memory Pane:} Represents a 2D list with individual list rows (\texttt{lst0}, \texttt{lst1}, etc.), and tracks both loop indices (\texttt{i}, \texttt{j}) as well as the current \texttt{total}.
    \item \textbf{Analogy Pane:} Uses a real-world metaphor of children’s gift wishlists and total spending, where each row represents a child’s list and items are tagged with prices.
\end{itemize}

\paragraph{Design Justification.}
\begin{itemize}
    \item \textbf{Wishlists as 2D Lists:} The metaphor supports the mental model of indexed access within nested collections, mapping intuitively to list-of-lists.
    \item \textbf{Visual Anchoring:} Highlights on the specific child and gift being accessed are synchronized with the corresponding memory cells and code line, facilitating index tracing.
    \item \textbf{Calculator Total:} The running total is abstracted as a visible cash register total, making the iterative accumulation concrete and reinforcing causal mapping.
\end{itemize}

\paragraph{Pedagogical Trade-offs.}
\begin{itemize}
    \item \textbf{Realism vs. Generality:} The toy store metaphor is highly relatable but domain-specific, and might obscure generalizable data access patterns (e.g., matrices or jagged arrays).
    \item \textbf{Memory Simplification:} Abstracts away list object references and memory model complexity in favor of schematic clarity, appropriate for novices but limiting for advanced learners.
\end{itemize}

\noindent
This visualization enables learners to grasp nested iteration and multidimensional indexing through layered metaphors and synchronized memory tracing, facilitating transfer to more complex data structure reasoning.

\subsection{Concept: While Loops and Conditional Search Termination}

\begin{figure}[h]
  \centering
  \includegraphics[width=0.95\linewidth]{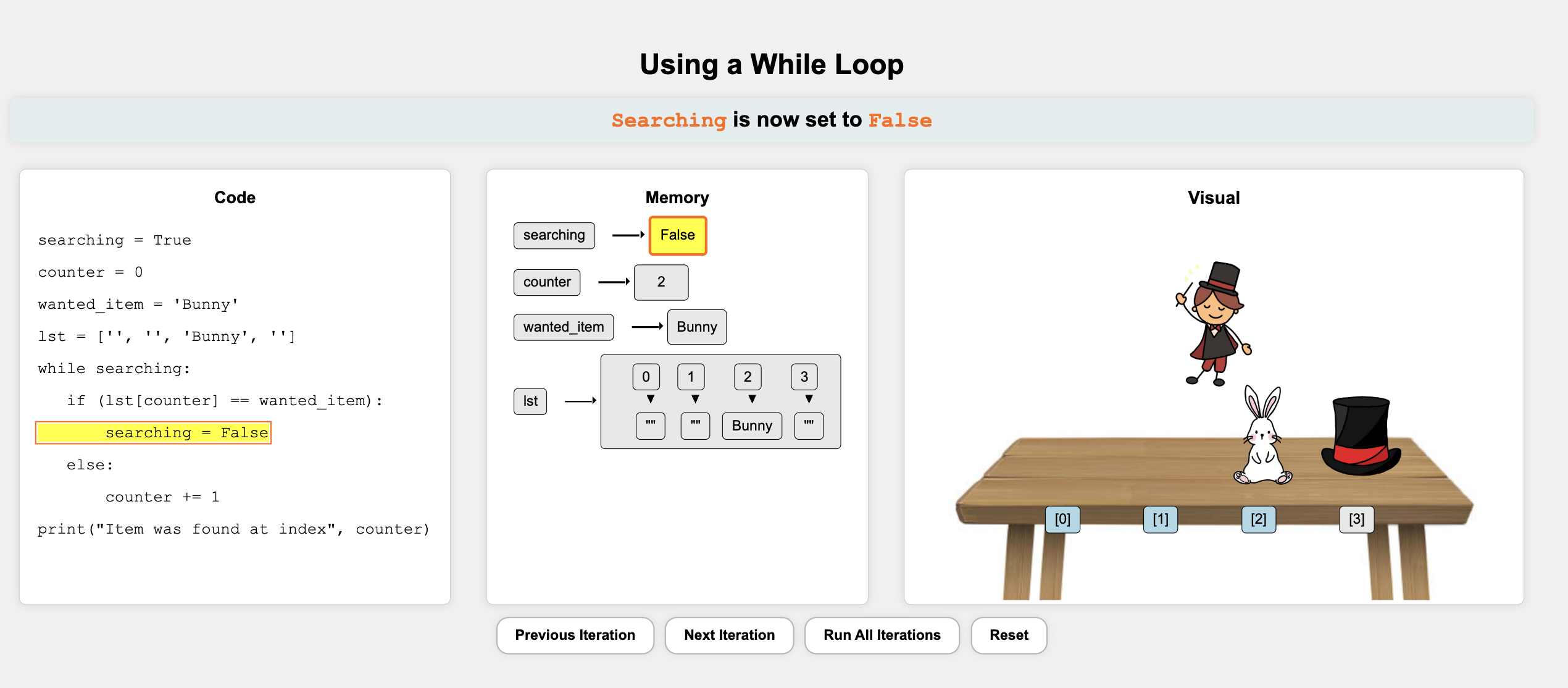}
  \caption{Visualization of a \texttt{while} loop searching a list until a target is found, modeled with a magician-and-hats metaphor.}
  \label{fig:while-search-visual}
\end{figure}

\paragraph{Learning Objective.}
To support understanding of loop control using sentinel variables, conditional branching inside loops, and early termination patterns when a match is found.

\paragraph{Representational Mapping.}
\begin{itemize}
    \item \textbf{Code Pane:} Emphasizes the loop condition (\texttt{while searching}) and its update logic (\texttt{searching = False}) inside an \texttt{if-else} structure.
    \item \textbf{Memory Pane:} Shows the current values of \texttt{searching}, \texttt{counter}, and the scanned list element (\texttt{lst[counter]}), as well as the static target value.
    \item \textbf{Analogy Pane:} Uses a magician metaphor where each list index is represented as a hat on a table, and the search is visualized as sequentially lifting hats to find a hidden bunny.
\end{itemize}

\paragraph{Design Justification.}
\begin{itemize}
    \item \textbf{Search Metaphor:} The magician uncovering hats mirrors the loop’s sequential search, providing a narrative that aligns well with procedural reasoning.
    \item \textbf{State Visualization:} The binary status of the \texttt{searching} variable is visually and textually reinforced to signal when and why the loop exits.
    \item \textbf{Synchronous Highlighting:} The active index and the comparison result are highlighted across all panes to support condition tracing and loop termination logic.
\end{itemize}

\paragraph{Pedagogical Trade-offs.}
\begin{itemize}
    \item \textbf{Narrative Specificity:} While charming and memorable, the magician metaphor may limit transferability to non-search while-loop patterns.
    \item \textbf{Boolean Logic Simplification:} The visualization avoids explicit discussion of Boolean evaluation expressions, which may reduce precision in learners’ understanding of loop guards.
\end{itemize}

\noindent
This visualization introduces learners to control structures that use sentinel variables and conditionally driven loop exit logic, grounded in a vivid search metaphor to make the abstract more concrete.

\subsection{Concept: Dictionaries and Key-Based Updates}

\begin{figure}[h]
  \centering
  \includegraphics[width=0.95\linewidth]{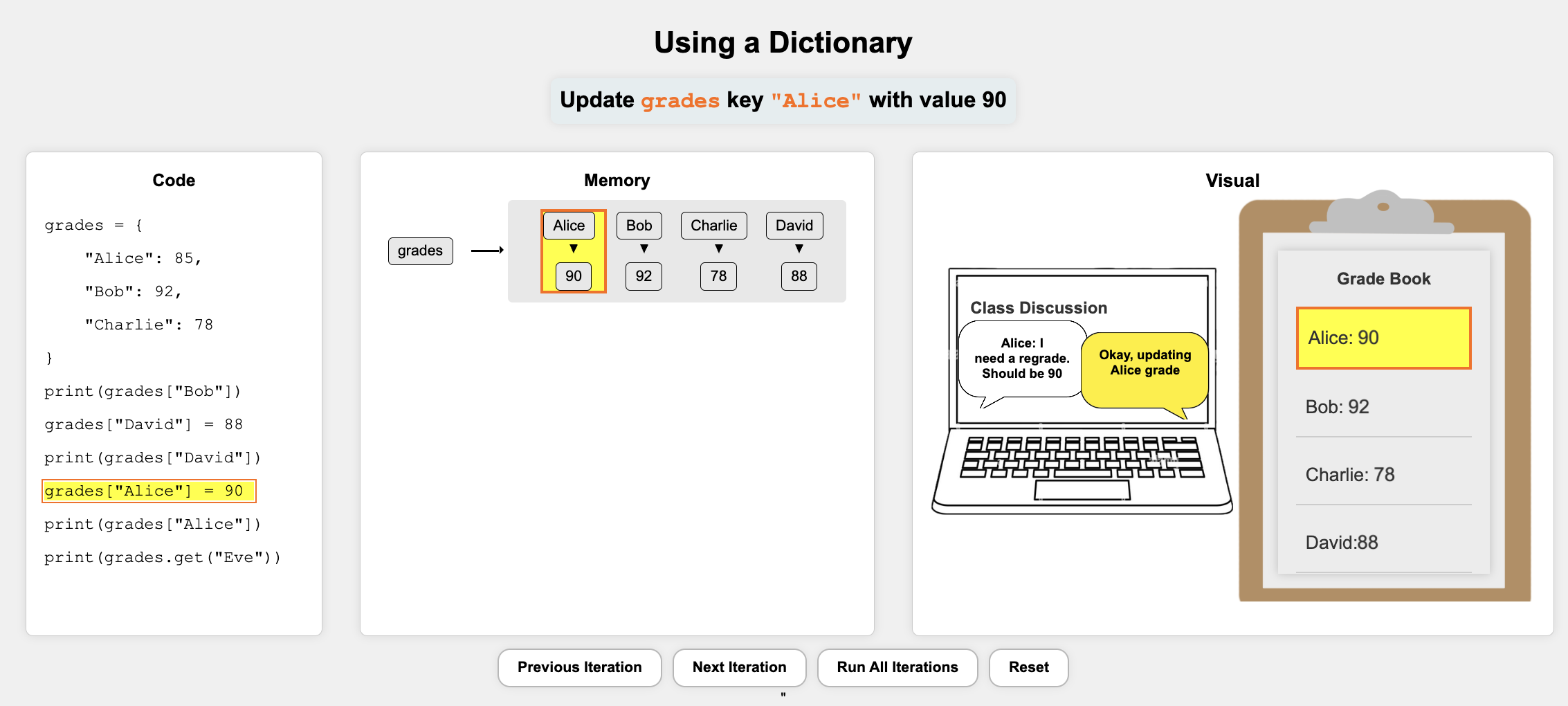}
  \caption{Visualization of dictionary update logic using a class-gradebook metaphor.}
  \label{fig:dict-update-visual}
\end{figure}

\paragraph{Learning Objective.}
To illustrate how dictionary structures associate keys with values and how those associations can be updated, retrieved, or queried safely.

\paragraph{Representational Mapping.}
\begin{itemize}
    \item \textbf{Code Pane:} Highlights dictionary creation, value retrieval using square-bracket indexing, and key-based updating.
    \item \textbf{Memory Pane:} Displays the dictionary as a key-value mapping, emphasizing the updated value associated with a specific key ("Alice").
    \item \textbf{Analogy Pane:} Uses a gradebook metaphor, where student names are keys and their scores are values. A laptop shows a conversational prompt representing a real-world rationale for an update (e.g., grade correction).
\end{itemize}

\paragraph{Design Justification.}
\begin{itemize}
    \item \textbf{Gradebook Metaphor:} This metaphor grounds the abstract concept of associative arrays in a familiar and practical context for students (especially in an academic setting).
    \item \textbf{Temporal Sequencing:} The "class discussion" visualization serves as a narrative anchor that justifies the update step, reinforcing comprehension of why and how the dictionary is modified.
    \item \textbf{State Highlighting:} The updated key is visually emphasized across all views (code, memory, analogy), helping learners attend to the causal link between the code statement and memory change.
\end{itemize}

\paragraph{Pedagogical Trade-offs.}
\begin{itemize}
    \item \textbf{Context Specificity:} The metaphor may be less effective in domains where dictionaries serve different functions (e.g., configuration objects or nested structures).
    \item \textbf{Overpersonalization Risk:} The dialogue balloon ("Alice: I need a regrade") may overly anthropomorphize the logic and obscure the general principle of key-value binding for some learners.
\end{itemize}

\noindent
This visualization helps demystify Python dictionaries by mapping updates to a well-scaffolded, socially contextualized interaction, supporting both declarative understanding and procedural reasoning.

\subsection{Concept: Sorting with Bubble Sort Algorithm}

\begin{figure}[h]
  \centering
  \includegraphics[width=0.95\linewidth]{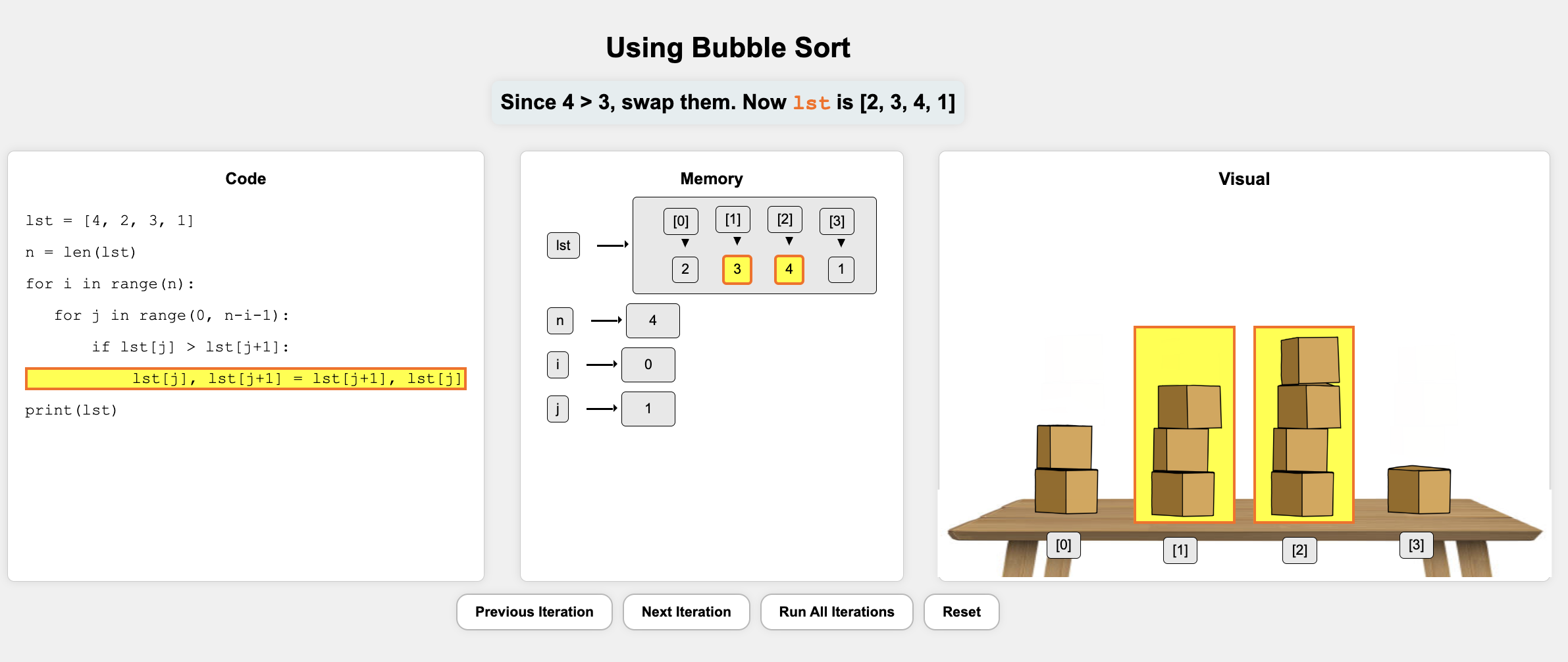}
  \caption{Bubble sort visualization using box-stacking to illustrate adjacent swaps.}
  \label{fig:bubble-sort-visual}
\end{figure}

\paragraph{Learning Objective.}
To develop an understanding of the iterative and comparative nature of the Bubble Sort algorithm, particularly its element-wise swapping behavior and loop nesting.

\paragraph{Representational Mapping.}
\begin{itemize}
    \item \textbf{Code Pane:} Highlights nested loops, pairwise comparison logic, and swap operation.
    \item \textbf{Memory Pane:} Displays the current list state and index markers \texttt{i}, \texttt{j}, and \texttt{n}, showing progression across iterations.
    \item \textbf{Analogy Pane:} Uses a tabletop of stackable boxes. Higher-valued boxes are visually taller, making it intuitive to detect which pair is out of order and should be swapped.
\end{itemize}

\paragraph{Design Justification.}
\begin{itemize}
    \item \textbf{Box Metaphor:} Height implicitly encodes value, making larger elements visually heavier or taller. This supports perceptual grounding of the "bubble" action as higher items moving upward through swaps.
    \item \textbf{Swap Visibility:} Highlighted comparison pairs and post-swap states emphasize the dynamics of local sorting rather than simply showing final outputs.
    \item \textbf{Memory-State Synchronization:} Visual and memory panes update simultaneously, reinforcing procedural understanding of index-based traversal.
\end{itemize}

\paragraph{Pedagogical Trade-offs.}
\begin{itemize}
    \item \textbf{Metaphoric Rigidity:} While stacking is intuitive for simple comparisons, it may not scale well to sorting algorithms involving complex heuristics (e.g., merge sort).
    \item \textbf{Loop Nesting Opacity:} Learners may not fully grasp the outer-inner loop interaction from a single iteration view; thus, additional timeline or iteration tracker may be needed.
\end{itemize}

\noindent
This visualization facilitates trace-based comprehension of Bubble Sort by anchoring comparisons and swaps in a tangible, object-based metaphor while maintaining fidelity to algorithmic structure.

\end{document}